\documentclass[article,amsart,fleqn,amsmath,floats,nofootinbib,aps,twocolumn,pra,floatfix,showpacs,superscriptaddress]{revtex4-2}

\usepackage{cmap}
\usepackage{wrapfig} 
\usepackage[pdftex]{graphicx}
\usepackage{pgfplotstable}
\usepackage[pdftex,unicode,colorlinks,bookmarks=false,citecolor=blue,linkcolor=blue,urlcolor=blue]{hyperref} 
\usepackage[T1,T2A]{fontenc} 
\usepackage{latexsym}
\usepackage{amsmath}
\usepackage{amssymb}
\usepackage{amsfonts}
\usepackage{color}
\usepackage{bm}
\usepackage{verbatim}
\usepackage[english]{babel}
\usepackage[utf8]{inputenc}
\usepackage{euscript}
\usepackage{subfigure}
\usepackage[normalem]{ulem}
\usepackage{listings}
\hypersetup{
colorlinks=true,
linkcolor=blue,
filecolor=magenta,
urlcolor=cyan,
}

\begin{document}


\title{Spectroscopy of electron-phonon interaction of superconducting point contacts: experimental aspects}


\author{N.~L.~Bobrov}
\affiliation{B Verkin Institute for Low Temperature Physics and Engineering of the National Academy of Sciences of Ukraine\\ prosp. Nauki 47, 61103 Kharkiv, Ukraine}

\email{bobrov@ilt.kharkov.ua}


\published {\href{https://doi.org/10.3367/UFNr.2019.11.038693}{\emph{UFNr}, \textbf{190}(11), 1143, (2020)};
[\href{https://doi.org/10.3367/UFNe.2019.11.038693}{\emph{UFNe}, \textbf{63}(11), 1072, (2020)}]}
\date{\today}

\begin{abstract}\emph{The recovering procedure of the electron-phonon interaction (EPI) functions from the additional nonlinearities of the current­voltage curve ($I-V$ curve) of point contacts associated with an excess current is considered. The approach proposed takes into account both inelastic scattering, which causes suppression of the excess current in the reabsorption of nonequilibrium phonons by electrons undergoing Andreev reflection (Andreev electrons), and elastic processes associated with the electron-phonon renormalization of the energy spectrum in a superconductor. The results obtained are systematically expounded for both the ballistic contacts, wherein the second derivatives of the $I-V$ curve in the normal state are proportional to the EPI functions, and inhomogeneous contacts (with dirty constrictions and clean banks), whose second derivatives in the normal state are either free of phonon singularities or weakly pronounced.}\\

\end{abstract}

\maketitle

\tableofcontents{}

\maketitle

\section{INTRODUCTION}
The first study \cite{1} of nonlinearities of the current-voltage ($I-V$) curve of point contacts with direct conductivity in the normal state employed as a source of information on the electron-phonon interaction (EPI), which laid the ground­work for Yanson's point-contact spectroscopy, was published in 1974. The fundamental theoretical study \cite{2} appeared only three years later, in 1977. The number of publications on Yanson's point-contact spectroscopy now exceeds several hundred; two monographs summarizing the results obtained have been published \cite{3,4}.

V.A.~Khlus and A.N.~Omel'yanchuk \cite{5,6} were the first to theoretically substantiate, as early as the 1980s, the feasibility of using the nonlinearity of the excess current (i.e., the difference between the $I-V$ curve for the superconducting and normal states of a metal in a contact under the same voltage) to reconstruct the EPI function. A theoretical justification for the same point contacts, but with a pre­dominant elastic component of electron-phonon scattering, was given in the late 1980s by A.N.~Omel'yanchuk, S.I.~Beloborod'ko, and I.O.~Kulik \cite{7}. Experimental studies, which showed the transformation of point-contact spectra in the transition of contacts to the superconducting state, appeared almost at the same time (see, for example, \cite{8,9,10,12,13,14,15,16}). However, the first experimental study in which the EPI function was reconstructed based on the difference between the second derivatives of the I-V curve of a point contact in the superconducting and normal states (below, for brevity, using the superconducting addition to the spectrum) was published almost 30 years later, in 2012 \cite{17}. The EPI function was first reconstructed, based on a superconducting addition to the spectrum, for point-contacts of superconductors, in which the elastic component of the current predominates,in 2014 \cite{18}. Published in the same year was study \cite{19}, in which an alternative method was applied to reconstruct the EPI function from the spectra of point contacts with excess current; it consists of a numerical solution of the Eliashberg equations using the McMillan-Rowell method \cite{20}.

The reason for such a time lag in the reconstruction of the EPI function from the superconducting nonlinearity of the $I-V$ curve is that, contrary to the predictions of the theory, neither the first derivative of the excess current nor the difference between the first derivatives of the $I-V$ curve of a point contact in the superconducting and normal states is proportional to the EPI function. Due to the presence of a superconducting background, which is not taken into account by the theory, it turns out that to reconstruct the EPI function the difference between the second derivatives of the $I-V$ curve should be used. Progress in this area was unencouraged by the theoretically predicted smallness of the additional nonlinearity of the $I-V$ curve in comparison with that of the normal state, which arises when a ballistic point contact undergoes transit to a superconducting state in the phonon energy range. The reconstruction of the EPI function based on this nonlinearity was of no practical interest, since superconductors with high critical parameters, whose transition to the normal state at low temperatures is either hindered or even impossible, proved to be beyond the scope of research. The way out of this situation was to use inhomogeneous point contacts, whose nonlinearity in the superconducting state significantly deviates from predictions of the theory. To reconstruct the EPI function with their help, the second derivative of the $I-V$ curve in the superconducting state at low temperatures suffices. A nice bonus is that such point contacts may be created more easily than ballistic ones.

We dwell on the elastic contribution in more detail in Section 3, while here we focus on inelastic superconducting EPI spectroscopy.
\section{Theory}

The simplest model of a point contact is two metal half-spaces separated by an infinitely thin spacer with a round hole of diameter $d$. The metals are continuously exchanging electrons through this hole. If the contact is small enough and operates in the spectroscopic regime (i.e., it is ballistic or diffusive), carriers are duplicated in the current-conducting state; electrons (holes) are divided into two groups, in which the difference between the energies of occupied and unoccupied electron states on the Fermi surface is $eV$, i.e., equal to the applied voltage \cite{21}. The electron can lose this excess energy at any point of its trajectory by emitting a nonequilibrium phonon. The mean free path of an electron with excess energy, referred to as the energy relaxation length (or inelastic mean free path), greatly depends on the value of this energy, reaching a minimum at energies equal to or exceeding the boundary energy of the phonon spectrum. The energy relaxation length in the spectroscopic regime is much larger than the contact diameter. Thus, most of the nonequilibrium phonons are generated far from the center of the contact, at its banks. Nevertheless, the highest density of nonequilibrium phonons is observed in the center of the contact, the region of the highest current density, and rapidly decreases as the current spreads. Therefore, in the first approximation, the point contact can be considered at large distances \textbf{r} from the constriction as a point source of phonons, whose density decreases as $\sim 1/\textbf{r}^2$ .

If the $I-V$ curve nonlinearities associated with phonons alone are taken into account, the following formula is valid for large biases at the contact:
\begin{equation}
\label{eq__1}
{I(V)=\frac{V}{{{R}_{0}}}+\delta I_{ph}^{N}(V)+I_{exc}^{0}+\delta I_{ph}^{S}(V)\ .}
\end{equation}

Here, ${{R}_{0}}$ is the resistance of the point contact in the normal state at zero voltage (this resistance is usually determined in experiments at a bias of several millivolts, and if the contact cannot be converted to the normal state, a good approximation is the differential resistance at $eV\gg\Delta$ (using the slope of a straight line that passes through zero parallel to the $I-V$ curve, under the appropriate bias)), $I_{exc}^{0}$ is the excess current independent of the bias at $eV\gg\Delta$, $\delta I_{ph}^{S}(V)$ is the negative addition to the excess current due to scattering of nonequilibrium phonons by Andreev electrons, and $\delta I_{ph}^{N}(V)$ is the negative addition to the current due to \emph{backscattering} of electrons in the contact region on nonequilibrium phonons.

\emph{Backscattering} processes are scattering events in which the electron returns to the same half-space which it left after an inelastic collision with a phonon. Due to geometric constraints, \emph{backscattering} processes are apparently only effective in a volume located sufficiently close to the constrict. Indeed, since the electron after scattering on a phonon can move in the isotropic case in any direction, the probability of such processes is at its maximum in the plane of the hole and equal to 0.5. As the distance from the center of the contact increases, the probability of backscattering rapidly decreases, and at quite large distances \textbf{\emph{r}} does not exceed the ratio of the contact area to the surface area of a sphere centered at the scattering point with radius \textbf{\emph{r}}, i.e., is $\sim(d/4r)^{2}$. Consequently, this probability at a distance $r\sim2d$ on the contact axis is $\sim32$ times less than at the contact center. If the scattering point is located away from the contact axis, the hole in the spacer looks like an ellipse, and the probability is correspondingly even lower.

Thus, most of the backscattering processes occur in a volume whose diameter is of the order of the point-contact diameter. Scattering on phonons at the contact banks, which make up most of the processes of scattering of electrons by phonons, is neither inverse nor makes any contribution to the nonlinearity of the $I-V$ curve.

Backscattering processes underlie Yanson's point-contact spectroscopy and, as shown in \cite{2}, the second derivative of this negative addition to the current is proportional to the EPI function (for a ballistic contact in the form of a circular hole with diameter $d$ and a spherical Fermi surface):
\begin{equation}
\label{eq__2}
{\frac{{{d}^{2}}\delta I_{ph}^{N}(V)}{d{{V}^{2}}}=-\frac{32ed}{3{{R}_{0}}\hbar {{v}_{F}}}G_{{}}^{N}\left( eV \right)}
\end{equation}

\begin{figure}[]
\includegraphics[width=8.5cm,angle=0]{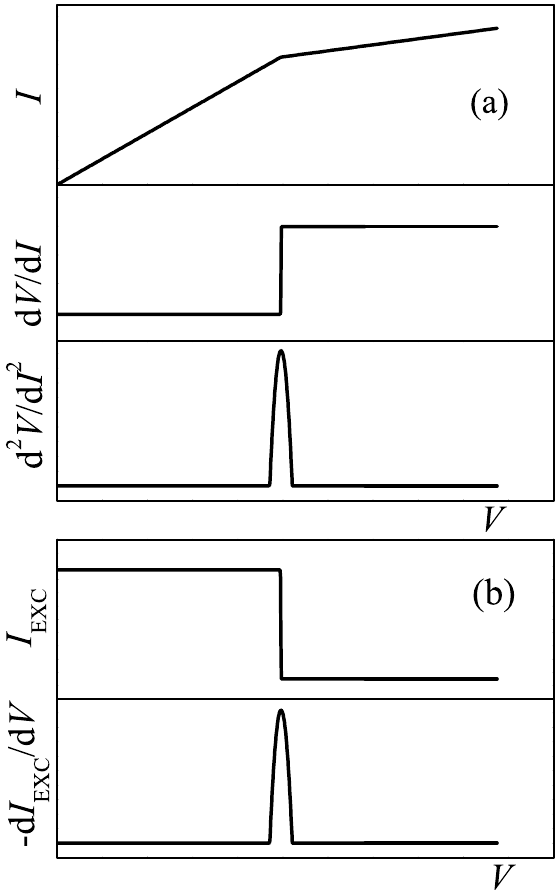}
\caption[]{(a) Manifestation of a single phonon line of an $N-c-N$ point contact on the $I-V$ curve and its derivatives. (b) Manifestation of a single phonon line of an $S-c-N$ and $S-c-S$ point contact in the excess current and its derivative. (See the description in the text.) }
\label{Fig1}
\end{figure}

Here, $G_{{}}^{N}\left( eV \right)$ is the EPI point contact function and $V_{F}$ is the Fermi velocity. Figure~\hyperref[Fig1]{1(a)} provides an easily comprehensible explanation as to why it is the second derivative of the $I-V$ curve that is proportional to the EPI function.
We assume that the phonon spectrum consists of a single line. It will manifest itself on the $I-V$ curve as a kink, which will look like a step in the first derivative of the $I-V$ curve, while in the second derivative it will consequently have the form of a peak. Smearing of curves is implied, owing to which the peak width and height are finite. Although theory \cite{2} involves $-d^{2}I/dV^2$, the figure displays the value $d^{2}V/dI^2$, which is usually recorded in experiments, since technically it is easier. These derivatives differ little in shape and are easily recalculated into each other using the simplest formulas. The magnitude of the nonlinear deviation of the resistance of the point contact from $R_0$ for ballistic contacts, which depends on the EPI strength, is several percent, but in any case is less than 10\%. Actual experimental curves differ from the schematic ones shown in Fig.\hyperref[Fig1]{1a}. Due to the low nonlinearity, the experimental $I-V$ curves are not very informative, since they differ insignificantly from a straight line. A typical first derivative of the $I-V$ curve of a ballistic point contact is displayed in Fig.\hyperref[Fig7]{7a} in Subsection~4.2, and the second derivative associated with it is shown in Fig.\hyperref[Fig5]{5a}. Seen in the first derivative are sections of the fastest growth, rather than steps, which correspond to peaks in the second derivative.

Let us consider the remaining terms of Eqn.\hyperref[eq__1]{(1)}. The excess current $I_{exc}^{0}$ does not depend on the bias and is proportional to $\Delta$ if $eV\gg \Delta$. The proportionality coefficient depends on the regime of passage of electrons through the hole. For ballistic $S-c-S$ contacts (superconductor-constriction-superconductor),
\begin{equation}
\label{eq__3}
{I_{exc}^{0}=\frac{8\Delta }{3e{{R}_{0}}}th\frac{eV}{2T}\ ,}
\end{equation}
i.e., if the inequality $d\ll \zeta$ , ${{v}_{F}}/{{\omega }_{D}}$, is satisfied; here, $d$ is the diameter of the contact, $\zeta$ is the reduced coherence length, $\tfrac{1}{\zeta }=\tfrac{1}{{{\xi }_{0}}}+\tfrac{1}{{{l}_{i}}}$,   $\xi_0$ is the superconducting coherence length, $l_i$ is the scattering length on impurities, and , ${{l}_{\varepsilon }}\sim {{{v}_{F}}}/{{{\omega }_{D}}}\;$ is the energy mean free path at the Debye energy. For diffusive $S-c-S$ contacts,
\begin{equation}
\label{eq__4}
{I_{exc}^{0}=\frac{\Delta }{e{{R}_{0}}}[\frac{{{\pi }^{2}}}{4}-1]th\frac{eV}{2T}\ .}
\end{equation}
Thus, the excess current in the diffusive mode is 55\% of that in the ballistic mode. The corresponding values for $S-c-N$ (superconductor-constriction-normal metal) contacts are half as much.

The term $\delta I_{ph}^{S}(V)$ in Eqn.~\hyperref[eq__1]{(1)} is a negative addition to the excess current due to the scattering of nonequilibrium phonons generated in the point contact by Andreev electrons. Then, already the first derivative of the excess current is proportional to the EPI function. Figure \hyperref[Fig1]{1(b)} shows the corresponding diagram. The phonon spectrum here is also represented by a single line, and the reabsorption of nonequilibrium phonons by Andreev electrons leads to a decrease in the number of these electrons and, consequently, to a stepwise decrease in the excess current. A peak will be observed in the first derivative of the excess current near the phonon energy. In other words, the phonon singularities in the first derivative of the excess current manifest themselves in the form of maxima of the differential resistance ($-dI/dV$).

It is of importance to highlight a few points in relation to these observations. Inasmuch as these scattering processes are not inverse, such scattering is effective at any distance from the hole where nonequilibrium phonons and Andreev
electrons coexist, since it results in a decrease in the excess current. However, the probability of such scattering is maximal, as in the case of inverse processes, near the hole in a volume whose diameter is of the order of the diameter of the contact, where the current density is still high. For the scattering to occur, the density of both phonons and Andreev electrons must be high; otherwise, they simply do not meet. The theory does not take into account scattering that occurs outside the diameter of the contact.

Thus, the areas of space in the ballistic contacts from which information about EPI comes for both mechanisms are very close, albeit not identical. Since the minimum size in which the value of the superconducting energy gap can vary coincides with the coherence length, while, according to the model conditions, it is much larger than the diameter of the point contact, such scattering processes do not lead to a change in the gap in the near-contact region. The suppression of excess current occurs here due to a slight decrease in the number of Andreev electrons.

The following was obtained in \cite{5} for $S-c-S$ contacts:
\begin{equation}
\label{eq__5}
{\frac{d{{I}_{exc}}}{dV}(V)\!=\!-\frac{64}{3R}\left(\frac{\Delta L}{\hbar \bar{v}} \right)\!{{\left[ {{G}^{N}}(\omega )\!+\!\frac{1}{4}{{G}^{S}}(\omega)\right]}_{\omega ={eV}/{\hbar }\;}}\ .}
\end{equation}
Here, $R$ is the resistance of the contact, $L$ is a function that has a very complicated form for arbitrary values of its arguments, $\bar{v}$ is the electron velocity averaged over the Fermi surface, ${{G}^{N}}(\omega )$ is the EPI point contact (PC) function, the same as in point contacts in the normal state in the Kulik-Omel'yanchuk-Shekhter (KOSh) theory \cite{2}, and ${{G}^{S}}(\omega )$ is the superconducting PC function of EPI that differs from ${{G}^{N}}(\omega )$ by the form factor. In contrast to the contribution to the current with a normal form factor due to backscattering, the contribution to the current in the case of a superconducting form factor contained in ${{G}^{S}}(\omega )$ is due to electron-phonon collisions associated with processes such as Andreev reflection in the contact area, i.e., with the transformation of quasielectronic excitations into quasi-hole ones.
A similar expression was derived in \cite{6} for the $S-c-N$ contact:
\begin{equation}
\label{eq__6}
\begin{matrix}
  \frac{1}{R(V)}-{{\left( \frac{1}{R(V)} \right)}_{\Delta =0}}= \\
  =-\frac{32}{3R}\cdot \frac{d\Delta }{\hbar }\cdot \left[ \frac{1}{v_{F}^{(1)}}\cdot {{G}_{1}}\left( \omega  \right)+\frac{1}{v_{F}^{(2)}}\cdot {{G}_{2}}\left( \omega  \right) \right] \ .\\
\end{matrix}
\end{equation}
Here, ${{G}_{i}}\left( \omega  \right)$, $i$=1,2 are the EPI function of the metals of which the contact consists. The relative value of the phonon contribution to the excess current at $eV\sim\omega_D$  is of the order of ${d\cdot {{\omega }_{D}}}/{{{v}_{F}}}\,$ i.e., is small if the condition ${d\ll {{{v}_{F}}}/{{{\omega }_{D}}}\;}$ is satisfied. This smallness is very important for the response of the excess current to the generated phonons to be linear, i.e., for the same change in the flux density of nonequilibrium phonons at different biases at the contact to cause the same change in the excess current. If the excess current is significantly suppressed by nonequilibrium phonons, as the bias at the contact increases, a similar change in the flux density of nonequilibrium phonons at high $eV$ will cause a smaller response. However, the suppression of the excess current may be caused by nonequilibrium phonons alone. Any mechanism that causes the excess current to be strongly suppressed will result in a violation of response linearity. To reconstruct the EPI function in such cases, the amplitude of the useful signal should be adjusted in the region wherein such a contribution is suppressed.

As noted above, the probability of reabsorption of nonequilibrium phonons by Andreev electrons for ballistic contacts, in which the condition $d\ll\zeta$ is satisfied rather accurately, is fairly high only near the constriction. As a result, the nonlinear deviation of the point-contact resistance from $R_0$, which is due to the scattering of nonequilibrium phonons by Andreev electrons, is several times less than the nonlinearity caused by backscattering processes. Therefore, if $eV\gg\Delta$, the spectrum changes for such contacts insignificantly in the transition to the superconducting state. Study \cite{6} describes such a transformation of the second derivative of the $I-V$ curve of the ballistic $S-c-N$ point contact:

\begin{equation}
\label{eq__7}
\frac{1}{R}\!\cdot \!\frac{dR}{dV}\!=\!\frac{16ed}{3\hbar }\!\cdot \!\sum\limits_{a=1,2}{\frac{1}{v_{F}^{(a)}}\!\cdot\! \int\limits_{0}^{\infty }{\frac{d\omega }{\Delta }\!\cdot \!S\left( \frac{\omega -eV}{\Delta } \right){{G}_{a}}(\omega )}\ .}
\end{equation}

where $G_{a}(\omega)$ is the EPI functions of normal and super­conducting metals that form a heterocontact, and $S(x)$ is the smearing factor,

\begin{equation}
\label{eq__8}
{S(x)=\theta (x-1)\frac{2{{\left( x-\sqrt{{{x}^{2}}-1} \right)}^{2}}}{\sqrt{{{x}^{2}}-1}},}
\end{equation}

with $\theta(y)$ being the Heaviside theta function. Thus, the spectrum is additionally smeared in the transition to the superconducting state, and, at $T\rightarrow0$, the resolution is determined by the value of $\Delta$.

Equation \hyperref[eq__7]{(7)}, taking into account the relationship between the derivative of the $I-V$ curve and the PC EPI function PC, yields

\begin{equation}
\label{eq__9}
\tilde{g}_{pc}^{S}(eV)=\int\limits_{0}^{\infty }{\frac{d\omega }{\Delta }S\left( \frac{\omega -eV}{\Delta } \right)g_{pc}^{N}(\omega )\ .}
\end{equation}

Thus, given the point contact spectra of a heterocontact in the normal and superconducting state, the results of calculations and experiment may be compared. In addition, as shown in \cite{18},

\begin{equation}
\label{eq__10}
g_{pc}^{S}(eV)=\frac{1}{\Delta }\int\limits_{0}^{eV}{\left[ \tilde{g}_{pc}^{S}(\omega )-g_{pc}^{N}(\omega ) \right]d\omega \ . }
\end{equation}

Equation \hyperref[eq__10]{(10)} apparently follows from the first derivative of the excess current being proportional to the EPI function.

\begin{figure}[]
\includegraphics[width=8.5cm,angle=0]{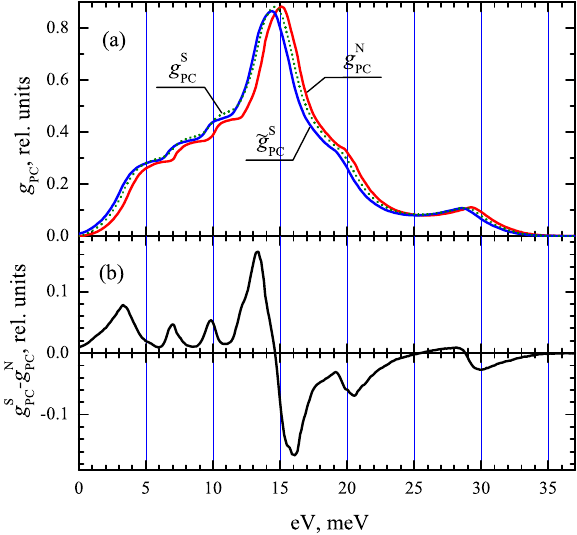}
\caption{(a) EPI function ${g}_{{pc}}^{{N}}$ $Sn-Cu$, reconstructed on the basis of the spectrum displayed in Fig.\hyperref[Fig3]{3a-d}.  ${\tilde{g}}_{{pc}}^{{S}}$ is the theoretically forecast transformation of the EPI function in transition to the superconducting state \hyperref[eq__9]{(9)} (see details in the text); ${g}_{pc}^{S}$ is the EPI function reconstructed based on the difference curve (b) by integration (Eqn \hyperref[eq__10]{(10)}). For easier comparison, the values of the ${g}_{pc}^{S}$ and ${g}_{pc}^{N}$ curves in the maximum are set the same.}
\label{Fig2}
\end{figure}

As can be seen, $\tilde{g}_{pc}^{S}(eV)$ reflects the transformation of the spectrum in the transition of the contact to the superconducting state. This function is proportional to the second derivative of the $I-V$ curve shifted to the lower-energy region by an amount of the order of $\Delta$ and additionally broadened, as a result of which it has a slightly lower intensity. The function $g_{pc}^{S}(eV)$ is proportional to the first derivative of the excess current, it does not contain additional broadening, and its shift to the lower-energy region is approximately one half as much.

\begin{figure*}[]
\includegraphics[width=16.5cm,angle=0]{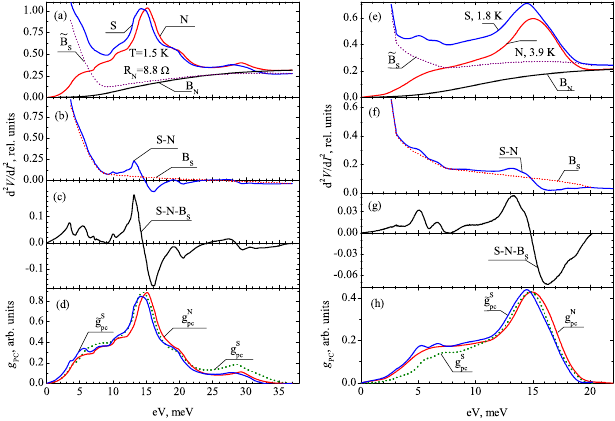}
\caption{Curves (a-d) for an $S-c-N$ $(Sn-Cu)$ point contact and (e-h) for an $S-c-S$ $(Sn-Sn)$ point contact. (a, e) EPI spectra of point contact in the normal $(N)$ and superconducting $(S)$ state, ${{\tilde{B}}_{S}}$ and $B_N$ are the background curves for the superconducting and normal spectra, respectively. (b, f) Difference between the superconducting and normal spectra and the assumed shape of the background curve. (c, g) Difference curve (after subtracting the background). (d, h) Point-contact EPI functions reconstructed from the normal ($g_{pc}^{N}$) and superconducting ($\tilde{g}_{pc}^{S}$) states and by integrating the difference curve ($g_{pc}^{S}$). Superconductivity is suppressed in the $S-c-N$ contact by the magnetic field, and in the $S-c-S$ contact by temperature.}
\label{Fig3}
\end{figure*}

Figure \hyperref[Fig2]{2} shows the EPI function reconstructed from the spectrum (Fig.\hyperref[Fig3]{3}) of the $Sn-Cu$ point contact in the normal state and the EPI functions calculated using Eqns \hyperref[eq__9]{(9)} and \hyperref[eq__10]{(10)}. The difference curve ${\tilde{g}_{pc}^{S}(\omega )-g_{pc}^{N}(\omega)}$ is also displayed.

\section{Production of contacts and measurement of their characteristics}

Mechanical clamping contacts between metal electrodes are the most widely used in point-contact spectroscopy. They may be both bulk electrodes made of high-purity metals and bulk electrode-film structures. A criterion for the purity of the material is the ratio $\rho_{300}/\rho_{res}$, where $\rho_{300}$ resistivity at room temperature and $\rho_{res}$ residual resistivity. Experience shows that, to obtain ballistic contacts, this ratio should be no less than $5\div10$, but usually materials with a ratio value of several dozen or higher are used. Electrodes may be any shape, although for convenience electrodes in the form of bars $10\div15$~mm long and with a cross section of $1\times1\div1.5\times1.5$~mm are most often used in studying pure metals. To eliminate mechanical hammer hardening, electrodes are made using an electroerosive machine, and the surface defect layer is removed by etching.

The selection of etching agents and regimes is critical for obtaining high quality point contacts. A thin nonconductive oxide layer should be formed on the surface of the electrodes that provides mechanical and electrical stability of the contact, since the area of the site where the electrodes touch each other is many orders of magnitude larger than that of the obtained contact. The contact is created by touching the edges of the electrodes and their subsequent shift relative to each other.

A high-purity metal wire is often used as one of the electrodes. Electrical molding of the contact is used in some cases, which consists of supplying a controlled, specially selected set of current pulses or just random external induction to a previously created high-resistance contact. In examining films deposited on a flat surface, a pointed needle or pyramid is often used as a counter electrode. The device for producing point contacts should be capable of both changing the force with which the electrodes are mutually pressed and displacing them relative to each other. To ensure the stability of the contact, one of the electrodes must be fixed to the damper.

In creating heterocontacts, copper and silver and, with some reservations, gold are most often used as the material of the counter electrode, since they can be easily processed and have many advantages over other conductors, which cannot be considered here in detail due to the limited volume of this review. The typical resistance of point contacts ranges from a few tenths of an ohm to several dozen ohms for metals with high conductivity and to several hundred ohms and higher values for low-conductivity compounds.

Apart from mechanical clamping contacts, a fairly wide range is available of point contacts that are created by other methods. For example, there are short-circuited film tunnel structures, break junctions, so-called soft touch contacts created using a tiny droplet of silver paste, and contacts created by the needle of a scanning tunneling microscope in short-circuit mode. Each of these methods has its own advantages for solving some specific problems, but also, accordingly, some disadvantages. There is no perfect uni­versal method. Mechanical clamping is the technique that is most widely used now to create ballistic contacts.

$I-V$ curve derivatives were measured using the standard modulation technique generally accepted in point-contact and tunneling spectroscopy. This technique is based on the following phenomenon: if apart from a constant current component a small alternating component is passed through a nonlinear element (point contact), the voltage across the element may be represented as a set of harmonics. The voltage of the first harmonic is proportional to the first derivative of the $I-V$ curve, that of the second harmonic is proportional to the second derivative, and so on. The effect of the modulating voltage is exhibited in the smearing of the spectrum, so to maintain a reasonable compromise between the resolution and the noise level, as low a voltage as possible is used.
\section{Ballistic point contacts}
\subsection{Contacts in which predictions of the theory are fulfilled}

We now consider the results of experiment \cite{18}. Figure\hyperref[Fig3]{3(a)} shows the spectra of the $Sn-Cu$ point contact in the normal and superconducting states, and the background curves for these spectra $B_N$ and ${{\tilde{B}}_{S}}$. Figure \hyperref[Fig3]{3b} shows the difference $S-N$ curve and the background curve $B_S$. The background is manifested in Yanson's PC spectroscopy as the difference between the second derivative of the $I-V$ curve and zero at biases larger than the limiting frequency of the phonon spectrum. At the same time, the EPI function vanishes. To subtract the background, an empirical self-consistent iterative procedure is most often used \cite{3}. It turns out that the background is also inherent in EPI superconducting spectroscopy: it is not sufficient to straightforwardly integrate the difference curve $(d^{2}{V}/dI^{2})_{S}-(d^{2}V/dI^{2})_N$ to obtain a curve proportional to the function EPI $g_{pc}^{S}(eV)$, since the second derivatives differ not only in the singularity associated with the gap at small biases, but also in the background level outside the phonon spectrum boundary. Because of this, the background curve BS should be subtracted from the difference curve prior to integration. In addition to the obvious requirement to be equal to zero outside the phonon spectrum, the difference curve after the superconducting background is subtracted must satisfy a sum rule: \textbf{the total areas under the curves above and below the abscissa axis must be the same.}

\begin{figure}[]
\includegraphics[width=8.5cm,angle=0]{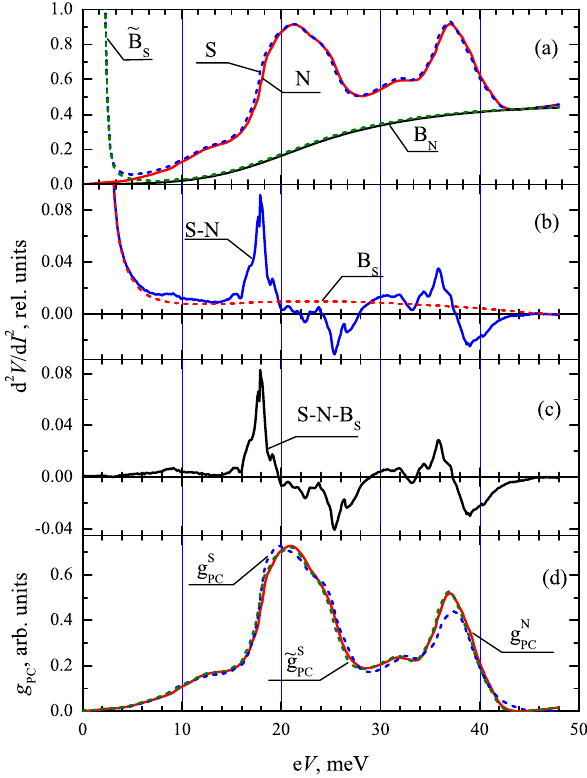}
\caption{(а) - (a) EPI spectra of an $Al-Al$ point contact in the normal and superconducting state $T/T_C$=0.68, $\Delta=0.85\Delta_0$. Superconductivity is suppressed by the magnetic field. (b) Difference between the superconducting and normal spectra and the assumed shape of the background curve. (c) Difference curve (after subtracting the background.) (d) Point­contact EPI function reconstructed by integration of the previous curve in comparison with the EPI function of the normal state.}
\label{Fig4}
\end{figure}

This requirement implies that we obtain after having integrated the difference curve a curve with a zero back­ground. Of course, many different curves $B_S$ can satisfy these criteria, but their variations, provided that the curves are smooth, do not result in noticeable changes in the shape and position of the phonon singularities of the EPI function under reconstruction.

And finally, ${{\tilde{B}}_{S}}$ is the background that must be extracted from the spectrum in the superconducting state to obtain a curve proportional to $\tilde{g}_{pc}^{S}(\omega )$. Given the spectrum in the normal state, the difference curve, and the background curves $B_N$ and $B_S$, it is easy to show that ${{\tilde{B}}_{S}}={{B}_{N}}+{{B}_{S}}$. Subtraction of the background from the spectrum in the superconducting state can apparently be replaced by adding two curves: the spectrum in the normal state with the
background $B_N$ subtracted and the difference curve with the background $B_S$ subtracted. The latter procedure is even more convenient to some extent, since it makes easier the approximation of the missing portion of the curve in the region of small biases near the gap singularity.

A comparison of Fig.\hyperref[Fig2]{2a} and \hyperref[Fig3]{3d} shows that there is good agreement between the predictions of the theory and the results of the experiment. In transiting to the superconducting state, the $\tilde{g}_{pc}^{S}(\omega )$ curves are smeared, decrease in amplitude, and shift to the region of lower energies by
order of the gap. The difference between the shapes of the experimental and theoretical curves $g_{pc}^{S}$ is observed in the high-energy region: the amplitude of the experimental curve is noticeably larger. This is apparently due to an increase in the density of nonequilibrium phonons at the periphery of the contact due to a decrease in the energy relaxation length of electrons at energies close to the Debye ones.

There are no formulas for $S-c-S$ point contacts similar to Eqns\hyperref[eq__7]{(7)} and \hyperref[eq__9]{(9)} for $S-c-N$ point contacts that describe the transformation of spectra in the transition from the normal state to the superconducting state. An indication alone is available in the case of $S-c-S$ point contacts that the positions of the phonon singularities on the EPI function reconstructed from the superconducting addition to the spectrum will coincide with those of the EPI function of the normal state.

Figures \hyperref[Fig3]{3e-h} show the results of an experiment to reconstruct the EPI function based on a superconducting addition to the spectrum for such a contact \cite{18}.
Although the temperatures of the normal and superconducting states for the $Sn-Sn$ point contact are not the same, in contrast to those for the $Sn-Cu$ point contact, the result of the reconstruction of the EPI functions is quite satisfactory.
The figure shows that in the transition of the contact to the superconducting state, no smearing of the spectrum is observed and, as a consequence, no decrease in its intensity, which occurs in the case of $S-c-N$ contacts. On the contrary, the intensity of the spectrum increases. As for the position of the phonon singularities, they also shift to the region of lower energies. While the shift of phonon maxima in the second derivative is of the same order as for the $S-c-N$ contacts, the shift for the EPI function reconstructed from the superconducting addition to the spectrum, albeit noticeably smaller, persists, in disagreement with the prediction of the theory.

\begin{table*}[htbp]
\caption[]{Characteristic parameters of tin and tantalum.}
\renewcommand{\tabcolsep}{0.25cm}
\begin{tabular}{|c|c|c|c|c|c|c|c|c|c|c|} \hline
& $R$, $\Omega$ & $d$, nm & $\rho_{300}/\rho_{res}$ & $\rho l$, $\Omega\cdot\text{cm}^2$ & $v_F$, cm/s & $l_{\varepsilon}^{D},$ nm & $l_i,$ nm & $\xi_0,$ nm & $\zeta,$ nm & $\Delta,$ meV\\ \hline
Sn & $7\div30$ & $5.1\div10.5$ & $\sim15000$& $4.5\cdot10^{12}$ & $1.89\cdot10^8$ &$\sim360$&6000&$\sim200$&$\sim200$&0.57 \\ \hline
Ta & $16\div210$&$2.2\div8.5$&18&	$5.9\cdot10^{-12}$&	$0.74\cdot10^8$&$\sim90$&	82&	92&	43&	0.71 \\ \hline
\end{tabular}
\label{Table1}
\end{table*}

It is noteworthy that the EPI function can be reconstructed based on the superconducting addition to the spectrum not only at $T\ll T_C$ (where $T_C$ is the temperature of the superconducting transition), as follows from the theory, but even at temperatures rather close to critical. Figure 5 in Ref.\cite{18} shows a set of EPI functions both obtained from the spectrum of the contact in the normal state and reconstructed from superconducting additions to the spectrum at various temperatures. Only slight variations in the shape of the curves are observed. The position of the main maximum remains unaltered even near the critical temperature, although the shape of this spectrum in the low-energy region already differs to some extent from the previous ones.

Figure\hyperref[Fig4]{4} shows the results for another metal, aluminum \cite{18}, which features a relatively low superconducting transition temperature and, consequently, small values of the gap and excess current. Therefore, the superconducting addition to the spectrum turns out to be very small: the curves $g_{pc}^{N}$ and $\tilde{g}_{pc}^{S}$ are virtually indistinguishable. Due to the inevitable errors associated with the digitization of the scanned experimental curves and their small difference, the accuracy with which the difference curve was determined turns out to be low, and therefore the shape of the EPI function reconstructed from this addition somewhat differs from that obtained from the spectrum in normal conditions. However, given the above circumstances, the agreement in shape may be considered quite satisfactory.
\subsection{Contacts in which the predictions of the theory fail}\label{1}

However, the agreement between theory and experiment demonstrated in Subsection 4.1 is not always the case, even for ballistic superconducting point contacts. Tantalum-based contacts are one such example \cite{22}. Deviations from theoretical predictions are observed for both heterocontacts and homocontacts. The strength of the effect, the magnitude of the superconducting addition to the spectrum, and the shape of the contribution differ. While the addition is relatively small for the previously considered point contacts, and the shape of the second derivative of the $I-V$ curve in the phonon frequency region changes very insignificantly in the transition to the superconducting state, the spectrum under­goes in the case of tantalum point contacts radical changes during the transition to the superconducting state. This is most clearly manifested for relatively low-resistance point contacts, but even for high-resistance point contacts, in which the ballistic condition is satisfied quite accurately, deviations from theoretical predictions are very large. No smearing of the spectrum and the associated decrease in intensity occur in the superconducting state. On the contrary, the phonon modes steeply sharpen. Moreover, the shape of the point contact spectrum changes, especially in the low-energy region. A soft phonon mode in the vicinity of an energy of $7\div8$~meV, which manifests itself in the normal state in the form of a plateau, is displayed in the superconducting state in the form of a peak; in addition, the amplitude of the phonon peak in the vicinity of an energy of $11.5$~meV significantly increases. However, the differences for the high-energy peak are noticeably less pronounced. To identify the reasons for such radical differences in the manifestation of phonon singularities in the spectra of tantalum-and tin-based contacts, we compare their characteristic parameters in Table\ref{Table1}.

\begin{figure}[]
\includegraphics[width=8.3cm,angle=0]{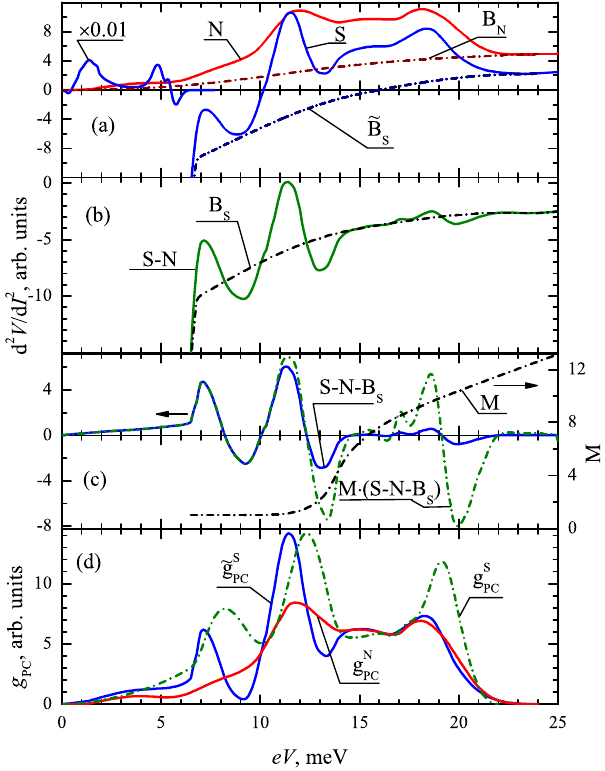}
\caption{(a) EPI spectra of a $Ta-Cu$ point contact in the normal and superconducting state. T=1.7~K, $R_0$=73~$\Omega$; the initial dashed segment of the superconducting curve that contains a gap and nonequilibrium singularity is scaled down by a factor of 100; ${{\tilde{B}}_{S}}$ and $B_N$ are the background curve for the superconducting and normal spectrum, respectively. (b) Difference between the superconducting and normal spectra and the assumed shape of the background curve. (c) Difference curve (after subtracting the background), scale curve $M$, and the difference curve multiplied by the scale curve. (d) Point-contact EPI functions reconstructed from spectra in the normal ($g_{pc}^{N}$) and superconducting ($\tilde{g}_{pc}^{S}$) states and from the superconducting addition to the spectrum ($g_{pc}^{S}$) by integrating the adjusted difference curve displayed in Fig. c. For easier comparison, the amplitudes of the curves $g_{pc}^{S}$ and the curve $\tilde{g}_{pc}^{S}$ are reduced to the same value. The scale of all curves is the same.}
\label{Fig5}
\end{figure}

Since the average energy relaxation length of electrons strongly depends on their energy, and the deviations in the tantalum spectra from the theoretical predictions in the transition to the superconducting state commence already in the region of small biases, the energy relaxation length is not significant for such differences. The main reason that results in the deviation of tantalum point contacts from the predictions of the theory is the small reduced coherence length $\zeta$. The volume of a sphere limited by the radius $\zeta$ is in tantalum 100~times less than the corresponding volume in tin; in addition, the Fermi velocity of electrons is also 2.5~times less. This leads to an increase in the density of Andreev electrons in the near-contact region and, consequently, increases the likelihood of their interaction with nonequilibrium phonons. Thus, contrary to the predictions of the theory, the reabsorption of nonequilibrium phonons by Andreev electrons occurs not only in a volume of the order of the diameter of the contact, but also in a volume whose characteristic size is the reduced coherence length.

In addition to phonon singularities in the transition to the superconducting state, nonspectral singularities emerge in the spectra of tantalum point contacts, the position of which depends on the temperature, resistance of the contact, and magnetic field. The existence of these singularities is also associated with the smaller reduced coherence length in tantalum compared to that in tin. These singularities are manifested in the form of a sharp surge in the second derivative of the $I-V$ curve, which is more intense than the gap singularity. This surge corresponds to a stepwise decrease in the excess current due to a sharp decrease in the superconducting gap in the near-contact region.
The reason for this behavior of the gap is that the critical density of nonequilibrium quasiparticles is attained in the near-contact region \cite{13,14,15}. Since the energy relaxation length in the contacts under consideration is much greater than their diameter, most of the electrons lose energy in the banks, being scattered on nonequilibrium phonons. These electrons are accumulated after the loss of excess energy above the gap, and their number increases with increasing voltage across the contact. When the critical density of nonequilibrium quasiparticles is attained, a jump-like transition of a part of the superconductor occurs to a new state with a partially suppressed gap. The smaller the superconductor volume adjacent to the constriction in which this transition occurs, the faster such a nonequilibrium state is apparently attained, i.e., the smaller the reduced coherence length that determines the minimum transition volume.

\begin{figure}[]
\includegraphics[width=8.3cm,angle=0]{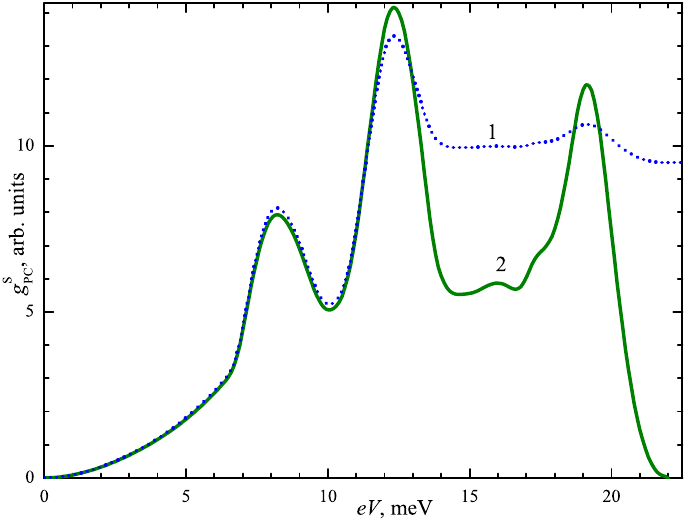}
\caption{Effect of the nonfulfillment of the sum rule on the shape of the reconstructed EPI function. The EPI function with the background (1) reconstructed from the curve ($S-N-B_S$) with an adjustment of the suppressed high-frequency part (Fig.\hyperref[Fig5]{5(c)}) and without the background (2) reconstructed from the adjusted curve M($S-N-B_S$). The scale of the curves is the same as in Fig.\hyperref[Fig5]{5}.}
\label{Fig6}
\end{figure}

\begin{figure*}[]
\includegraphics[width=16.5cm,angle=0]{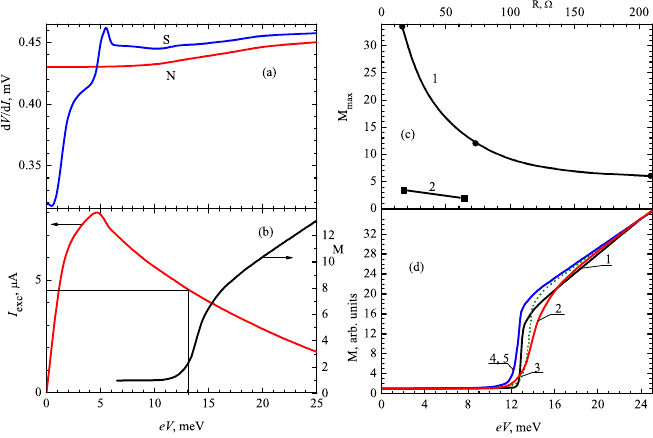}
\caption{(a) First derivatives of the $I-V$ curve for a point contact, the characteristics of which are displayed in Fig.\hyperref[Fig5]{5}, in the normal (N) and superconducting (S) states. (b) Excess current as a function of bias; also shown are the scale curve M (see Fig.\hyperref[Fig5]{5c}) and the values of the excess current and voltage that correspond to the onset of the growing section on the scale curve. (c) Value of the adjusting scale curve at the edge of the phonon spectrum at 23~meV as a function of resistance for heterocontacts (curve 1) and homocontacts (curve 2). (d) Adjusting scale curves for contacts of various resistances reduced to the same amplitude: 1~-~$Ta-Cu$, R=209~$\Omega$; 2~-~$Ta-Cu$, R=73~$\Omega$; 3~-~$Ta-Cu$, R=16~$\Omega$; 4~-~$Ta-Ta$, R=64~$\Omega$;and 5~-~$Ta-Ta$, R=17~$\Omega$. Curves 4 and 5 have the same shape, but differ in amplitude by a factor of approximately 2.7 (see Figs~12 and 14 in \cite{22}).}
\label{Fig7}
\end{figure*}

The position of the nonequilibrium singularity for Ta-Cu heterocontacts of various resistance values at a temperature of 2 K corresponds to an injection power of approximately 0.4~$\mu$W. An increase in the temperature or a magnetic field results in an increase in the relaxation rate of quasiparticles above the gap; therefore, to attain their critical density, the injection power should be increased. The resulting shift of the nonequilibrium singularity towards high voltages rules out the interpretation of such singularities as related to the breakup of superconductivity due to heating or suppression by a magnetic field.

Singularities of this kind are not observed in the spectra in the phonon frequency region for $Sn-Cu$ heterocontacts, since the volume in which a transition to a nonequilibrium state with a suppressed gap can occur is two orders of magnitude larger.

Since the transition to a nonequilibrium state quite insignificantly suppresses the excess current, the effect of this transition on the superconducting addition to the spectrum is also small, with the exception of a possible coincidence in energy with phonon singularities. The significantly higher intensity of the nonequilibrium singularity does not allow in this case the phonons to be traced. A further increase in the injection power with an increase in the voltage applied to the contact leads to suppression of the smoothness of the excess current. This suppression depends on the 8 resistance of the contact and can be very significant for low­resistance contacts, which noticeably affects the intensity of the singularities associated with the scattering of nonequilibrium phonons by Andreev electrons.

We now consider the specific features of the formation of phonon singularities in the spectra of tantalum-based point contacts in the superconducting state. The superconducting addition to the spectrum in such contacts, in contrast to that for the previously considered tin- and aluminum-based contacts, is a superposition of contributions from a region whose size is of the order of the diameter of the contact and a region with a characteristic size of the reduced coherence length, which is significantly larger in volume. Owing to the latter circumstance, phonons with low group velocities that correspond to the maxima of the phonon density of states start playing an essential role in the reabsorption of nonequilibrium phonons by Andreev electrons. Such phonons leave this volume more slowly, which leads to an increase in their density in comparison with that of fast phonons. Since the scattering probability depends on the phonon density, the contribution of slow phonons with energies corresponding to phonon peaks will be the largest --- the peaks will sharpen. Phonon selection is not operative in a region whose size is of the order of the diameter of the contact --- all phonons are equivalent. The ratio of the contributions from these volumes determines the degree to which the phonon peaks in the spectrum sharpen in the transition of the contact to the superconducting state. The degree of sharpening apparently increases with an increase in the size of the contact, an observation that is perfectly confirmed by experiment --- the sharpening is stronger for low-resistance contacts.

Prior to proceeding to the experimental results for tantalum contacts, it should be noted that their contribution is not manifested in the point-contact spectra of heterocontacts of transition \emph{d}-metals or compounds based on them and $Cu$, $Ag$, and $Au$. This circumstance is very convenient for studying the spectra of these metals or compounds based on them.

We now consider a rather high-resistance $Ta-Cu$ contact with a resistance of about 73~$\Omega$ (Fig.~\hyperref[Fig5]{5}). Its diameter is $\approx3.7$ nm, and its parameters fully comply with the requirements of the theory \cite{6} (see Table~\ref{Table1}); however, the transformation of the contact spectrum in the transition to the superconducting state does not agree with theoretical predictions. The nonequilibrium singularity in the spectrum in the superconducting state is located for this contact at an energy of about 5~meV, and its energy only coincides with that of the soft mode at an edge. The shape of the superconducting soft mode is a peak rather than a plateau, as in the normal state. Observed are a sharpening and a significant increase in its intensity rather than smearing of the phonon peak in the energy region of 11.3~meV. However, the effect of the superconducting transition on the peak in the 18~meV region is noticeably less. As can be seen in Fig.~\hyperref[Fig5]{(5b,c,)} the sum rule formulated in Subsection~4.1 is not satisfied for the resulting $S-N-B_S$ curve after the superconducting background $B_S$ is subtracted. Due to this, attempts to reconstruct the EPI function from this curve yield a $g_{pc}^S$ curve with a background. To correct this circumstance, the $S-N-B_S$ curve should be adjusted in the energy range above 12~meV. The adjustment is carried out by multiplying the $S-N-B_S$ curve by an empirical scale adjustment curve M. The curve M is 1 in the low energy region (does not change the shape of the difference curve) and increases the amplitude of the difference curve in the high energy region in such a way that the adjusted curve satisfies the sum rule. The results of reconstructing the EPI function from the resulting curves $S-N-B_S$ and $\text{M}(S-N-B_S$) are displayed in Fig.\hyperref[Fig6]{6}.

As noted above, the reason for the decrease in the amplitude of the resulting $S-N-B_S$ curve in the energy region above 12~meV is the suppression of the excess current by nonequilibrium quasiparticles. Figure\hyperref[Fig7]{7a,b} shows the dependences of the first derivatives of the $I-V$ curve of the point contact in the normal and superconducting states, the dependence of the excess current on the contact bias, and the scale adjustment curve $M$, the same as in Fig.\hyperref[Fig5]{5}. For easier comprehension, vertical and horizontal segments indicate the voltage and excess current values that correspond to the onset of the step on the scale curve. As can be seen from the figure, the dependence of the excess current on M in the vicinity of the step is free of singularities.

It should be noted that suppression of the superconducting addition to the spectrum is observed for all tantalum­based ballistic point contacts at biases above 12~meV. A consequence of this phenomenon is that the sum rule for the resulting curves $S-N-B_S$ is violated and scale adjustment curves need to be applied in reconstructing the EPI function. These curves proved to be characterized by almost the same same curve may be used for various contacts by tuning its amplitude, which is demonstrated using homocontacts as an example. At the same time, the amplitude of the curves very greatly depends on the degree of suppression of the excess current, which correlates with the resistance of the contact. The excess current for homocontacts with comparable resistances is twice as large; therefore, the amplitude of the scale curves is also lower. Although all scale curves exhibit a step in the 12$\div$13~meV region, the dependences of the excess current on the voltage for all the given contacts do not exhibit any clearly pronounced singularities in this range of biases; only a sufficiently smooth decrease in the excess current is observed. At the same time, a sharp decrease in the density of phonon states is observed near this energy for tantalum after the first peak. As noted above, the probability of scattering of nonequilibrium phonons by Andreev electrons in a volume limited by the reduced coherence length highly depends on the density of both components. It can be assumed by analogy with the mechanism of the emergence of nonequilibrium singularities that there are certain threshold densities of both Andreev electrons and nonequilibrium phonons --- the probability of their interaction sharply drops at lower densities. Moreover, these threshold densities are interrelated --- if the density of one component increases, the critical density of the other decreases. The initiating factor is in this case a sharp decrease in the rate of change in the flux density of nonequilibrium phonons at energies of about 12~meV against the background of a smooth decrease in the excess current. The magnitude of the effect in this case already depends on the degree of suppression of the excess current: as the previous analysis shows, the decrease in the probability of scattering (and, consequently, the magnitude of the superconducting addition to the spectrum) can change severalfold. This is also confirmed by homocontacts --- larger values of excess currents correlate with smaller amplitudes of the adjustment curves. As follows from the shape of the adjusting scale curves, further suppression of the excess current, as well as the rate of change in the flux density of nonequilibrium phonons, is no longer accompanied by any jumps.

We note in concluding this section that a stepwise decrease in the amplitude of the superconducting addition to the spectrum against the background of smooth suppression of the excess current is not only observed for tantalum-based contacts, but also characteristic of all superconducting contacts studied by us, which deviate from theoretical predictions.
\section{Inhomogeneous point contacts}

Ballistic contacts are very important for understanding the processes that occur in them in transiting to the super­conducting state and for clarifying the procedures needed to reconstruct the EPI function. Inhomogeneous contacts are a clue to reconstructing the EPI function in superconductors, the transition of which to the normal state at low temperatures and production of ballistic contacts based on them are very challenging tasks. An inhomogeneous point contact is one which contains in its constriction a region with a diffusive mode of electron motion, i.e., a region for which the inequality ${{l}_{i}}\ll d\ll {{\Lambda }_{\varepsilon }}$ is satisfied, where $d$ is the diameter of the contact, $l_i$ is the elastic relaxation length, $l_\varepsilon$ is the energy relaxation length, and ${{\Lambda }_{\varepsilon }}$ is the diffusive energy relaxation length, ${{\Lambda }_{\varepsilon }}=\sqrt{{{l}_{i}}{{l}_{\varepsilon }}/3}$. The intensity of the spectrum in the diffusive regime is an order of magnitude lower than in the ballistic regime by the factor $\sim l_{i}/d$.

According to theoretical model \cite{23}, the second derivative of the $I-V$ curve of the diffusive contact differs from that for the ballistic one by its smaller value. In addition, an almost complete isotropization of the EPI spectrum is observed, which is due to the isotropization of the electron momentum distribution. This can only lead to a slight smearing of the spectrum and an insignificant change in its shape.

For the contact to be a diffusive one, it is not required that the defect zone occupy the entire volume of the order of the diameter of the contact. If the highest density of defects and impurities is concentrated near the center of the contact and decreases at the periphery, the electrons --- after being scattered by nonequilibrium phonons --- diffuse with high probability in the direction in which the density decreases. In other words, the spectrum in the normal state is primarily determined by scattering processes near the center of the contact at the interface at which electrodes touch each other. This can lead to an increase in the contribution to the EPI spectrum of surface phonons and, consequently, to its additional smearing. If only one bank of the point contact is in the diffusive mode, the intensity of each partial contribution in the resulting spectrum is apparently determined by the dirty bank.

The diffusive regime is attained in experiments with a high density of defects and impurities in the constriction region. The diffusive condition for point contacts of relatively small diameters assumes that the elastic relaxation length is comparable by an order of magnitude to the period of the crystal lattice, which turns out to be highly distorted in the perturbed region. The second derivative of the $I-V$ curve of such contacts in the normal state exhibits strongly smeared low-intensity phonon singularities against a very high back­ground level or even their complete absence. The derivative actually becomes similar to a background curve.

Nevertheless, the regime of passage of electrons through the constrictions remains a spectroscopic one: carriers are duplicated in the current-passage state; the difference between the energies of occupied and free states of electrons in the current-passage state on the Fermi surface remains equal to the applied voltage, and weakly pronounced and smeared phonon singularities in the normal state remain in their places.

For spectroscopy in the superconducting state to be successful, it is also necessary that the point-contact banks adjacent to the constriction have a sufficiently perfect crystal lattice, since nonequilibrium phonons produced in the banks reflect the vibrational structure of the material in the region of its production. Since the quality of the material in the bulk of the electrode near the short circuit can alter in the process of manufacturing the contact, the obtained spectrum is the most reliable source of information about the quality of the lattice. For these phonons to effectively interact with Andreev electrons, the reduced coherence length in a superconductor should not be very large. These requirements are usually sufficient to reconstruct the EPI function based on the second derivative of the $I-V$ curve of a point contact in the superconducting state.
The nonlinearity of the $I-V$ curve in the superconducting state depends on both the magnitude of excess current and the volume in which the Andreev electrons effectively interact with the generated nonequilibrium phonons. Since this current in diffusive contacts is slightly more than half of the value in ballistic contacts and the effective volume increases due to the involvement of the banks at a distance of the order of the reduced coherence length, it may be expected that the nonlinearity of the $I-V$ curve of point contacts in the region of phonon frequencies in the superconducting state will be much larger than in the normal state. A similar situation is often realized in clamping point contacts, since the surface is usually more contaminated than the bulk of the material, and additional distortions are introduced at the interface between the electrodes in the process of creating a point contact.

\subsection{ Niobium­based point contacts}
It is reasonable to test the arguments set forth above on a well­studied superconductor. Niobium is a suitable candidate from this perspective. On the one hand, the EPI function for this metal has been well studied. On the other hand, niobium is a very complex metal from the perspective of point-contact spectroscopy. A large number of centers that scatter electrons are available on its oxidized surface, while they are absent in the bulk of the metal. A very large number of experiments have been carried out in studying EPI in niobium using Yanson spectroscopy. Several hundred contacts have been explored, the overwhelming majority of whose spectra correspond to the dirty limit. Only a few point-contact spectra in the normal state exhibited the presence of phonon singularities, and their quality was rather low \cite{24,25}. It is all the more interesting to confirm the feasibility of reconstructing the EPI function using inhomogeneous point contacts.

\begin{figure*}[]
\includegraphics[width=16.5cm,angle=0]{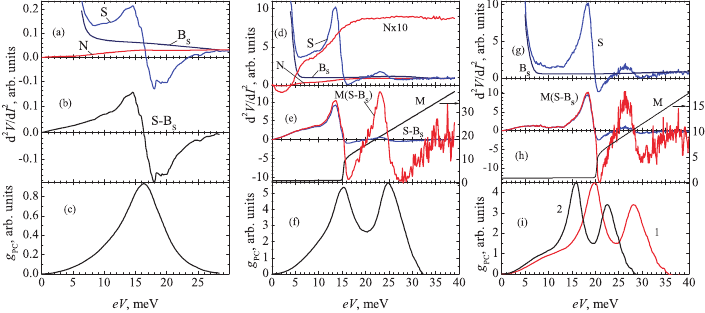}
\caption{Curves for inhomogeneous niobium type I (a-c) and type II (d-i) contacts. (a, d, g) EPI spectra of a point contact in the superconducting ($S$) and normal ($N$) state (the curve labelled $N\times 10$ shows the normal state spectrum magnified by a factor of 10), $B_S$ is the background curve for the superconducting spectrum, respectively. (b, e, h) Superconducting spectrum with the background subtracted. (c, f, i) Point-contact EPI functions reconstructed from the superconducting state by integrating the difference curve; in Fig. i, 1 is the initial curve and 2 is the adjusted curve (see details in the text.)}
\label{Fig8}
\end{figure*}

To create point contacts, quite pure niobium was used, in which the elastic relaxation length at a low temperature was $\sim$220~nm. The coherence length in niobium $\xi_{0}\sim44$~nm; given the elastic relaxation length, the reduced coherence length in an unperturbed material is $\zeta\sim36$~nm.

Two approaches were employed in creating the contacts. In the first one, the electrodes used to create contacts were etched in a mixture of acids, rinsed in water, and dried before being mounted in a clamping device. In the second approach, the electrodes treated by the above method were placed in an ultrahigh vacuum and heated to a pre-melting temperature using an electron beam. A layer of aluminum with a thickness of $\sim10\div15$~nm was deposited after cooling on the edge of the electrodes and next oxidized in a boiling 30\% solution of $H_{2}O_2$. In what follows, point contacts created using the electrodes produced within the first approach are called for brevity type I contacts, and those produced using the second approach, type II contacts.

The second derivatives of the $I-V$ curve of contacts of both types in the normal state were very similar and exhibited a high background level with an almost complete absence of phonon singularities. In the superconducting state at a constant level of modulating voltage, the value of the second derivatives of the $I-V$ curve increased by approximately an order of magnitude, while for contacts of different types they usually differed noticeably in shape. The most characteristic feature of type I contacts is an inverted $S$-shape without any additional structure, which can be seen in Fig.\hyperref[Fig8]{8а}. Since the nonlinearity of the $I-V$ curve in the superconducting state is primarily determined by the scattering of nonequilibrium phonons by Andreev electrons, an additional insignificant contribution from the backscattering can be ignored --- there is no need to subtract the second derivative of the $I-V$ curve in the normal state to reconstruct the EPI functions. However, the procedure for subtracting the background $B_S$, taking into account the fulfillment of the sum rule, apparently remains in force. After having integrated the difference curve (Fig.\hyperref[Fig8]{8b,e,h}), we obtain the EPI function (Fig.\hyperref[Fig8]{8c,f,i}). Since the superconducting addition to the spectrum is formed in a volume whose characteristic size is the length of conversion of Andreev electrons into Cooper pairs, the shape of the EPI function reconstructed based on this addition may be used to draw an indirect conclusion about the degree of perfection of the crystal lattice of this volume. As can be seen from the figure, the EPI function reconstructed for this contact is a smeared bell-shaped curve, typical of a highly deformed metal. A corresponding example is given in \cite{26} in Fig.~1 for the spectrum of a zirconium contact produced by the fracture method. However, a gentler technique of producing clamping contacts enables obtainment of zirconium spectra with a significantly better resolution, even if electrodes of apparently inferior quality are used (see Fig.~9 in Ref.\cite{27}).

Thus, the thickness of the defect layer in type I niobium contacts most often turns out to be comparable to the length of conversion of Andreev electrons into Cooper pairs or even to exceed it. Therefore, if nonuniform point contacts are used to reconstruct the EPI function, the key condition is the minimum possible thickness of the defect layer. At the very least, it must be much smaller than the size of the region in which nonlinearity is formed in the superconducting state. It turns out that this condition is much simpler to satisfy for type II contacts.

Figures\hyperref[Fig8]{8d-f} show the spectra of the type II $Nb-Nb$ point contact in the normal and superconducting states. It is difficult to estimate in quantitative terms the diameter of such a point contact, since the electronic parameters of the defective material, of which the constriction consists, are unknown. It is only possible to use the electronic parameters of pure niobium to obtain a lower bound for the diameter by means of the Sharvin formula. We get for a resistance of 17.5~$\Omega$ $d>6$~nm. In the normal state, there are no peaks in the spectrum in the region of the corresponding phonon energies; a small plateau alone is observed near the soft mode. The intensity of the normal spectrum is extremely low, which corresponds to a small elastic scattering length and is characteristic of an amorphous material. For better comprehensibility, the curve marked as N$\times$10 shows this spectrum magnified by a factor of 10. The intensity of the spectrum increases by an order of magnitude in the transition to the superconducting state, and singularities associated with phonons emerge in that spectrum. The shape of the spectrum as a whole is similar to that of the $S-N$ difference curve for tantalum point contacts (see, for example, Fig.\hyperref[Fig5]{5}), but with a number of significant differences. Non-equilibrium singularities observed for tantalum ballistic contacts are missing. Judging by the absence of a peak at the soft mode site, there is no selection of phonons with low group velocities. A probable explanation is that lattice defects can result in a small elastic free path of nonequilibrium phonons, which counterbalances the selection effect.

\begin{figure}[]
\includegraphics[width=8.3cm,angle=0]{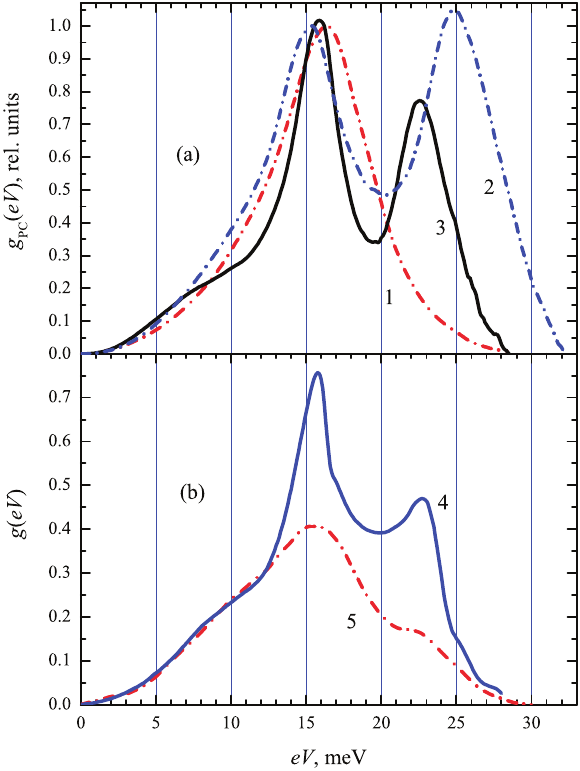}
\caption{(a) EPI functions (curves 1-3) of $Nb$ point functions normalized to one in the region of the first peak (see, respectively, the contacts displayed in Figs \hyperref[Fig8]{8a-c, 8d-f, and 8g-i}). (b) EPI function of Nb reconstructed from the tunneling spectrum in \cite{28} (curve 4) and the second derivative of the $I-V$ curve of the point contact in the normal state (curve 5) in \cite{24,25}.}
\label{Fig9}
\end{figure}

Figures \hyperref[Fig8]{8b, e, h} show the difference curve after subtracting the background $B_S$. As in the case of tantalum-based point contacts, the high-energy part of the difference curve is noticeably suppressed and requires adjustment. Figures \hyperref[Fig8]{8c, f, i} show the EPI function reconstructed using the adjusted curve. Despite the apparent progress compared to the previous contact, the quality of the reconstructed EPI function is still not high enough. Phonon peaks are smeared: while the first peak is located at approximately 16~meV, which corresponds to the position of the maximum on the EPI tunneling function, the high-frequency peak is shifted to the region of higher energies, from 22 to 25~meV. Such behavior previously observed for some spectra in the normal state \cite{24,25} was associated with inhomogeneity of the material in the region where the $I-V$ curve nonlinearity is formed, in particular, with the presence of impurities in the form of conducting niobium oxides.

\begin{figure*}[]
\includegraphics[width=16.5cm,angle=0]{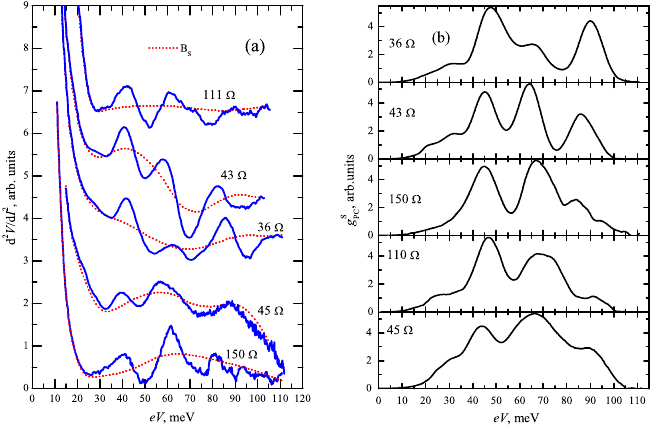}
\caption{(a) Point-contact spectra of $MgB_2-Ag$ heterocontacts (T=4.2~K, H=0) taken from \cite{29,30,31} and the assumed background curves. For easier comprehension, the curves are shifted vertically. (b) Point-contact EPI functions of $MgB_2$ reconstructed from difference curves.}
\label{Fig10}
\end{figure*}
\begin{figure}[]
\includegraphics[width=8.3cm,angle=0]{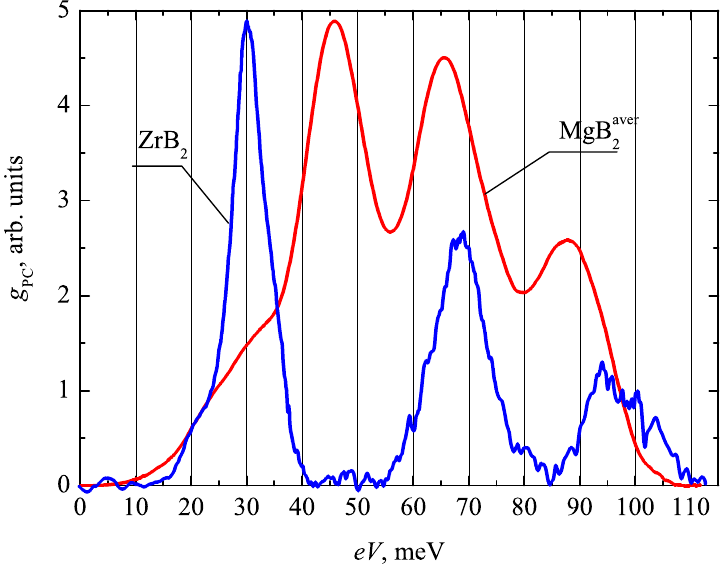}
\caption{Point-contact EPI function of $MgB_2$ averaged over the five spectra displayed in Fig.\hyperref[Fig10]{10} in comparison with the EPI function of $ZrB_2$ taken from \cite{30}.}
\label{Fig11}
\end{figure}

Of significant interest is the spectrum of the type II contact shown in Fig.\hyperref[Fig8]{8g-i}. Its shape is similar to that of the spectrum in Fig.\hyperref[Fig8]{8d-f}; however, all the singularities of the spectrum are shifted to the region of high energies. After subtraction of the background $B_S$, adjustment of the resulting curve, and subsequent integration, we obtain the EPI function, the peaks of which are located near 20 and 28~meV. This corresponds to the niobium EPI function "stretched" along the $x$-axis by 20\%. Multiplying the $x$-coordinate of the curve by 0.8, we obtain the EPI function, which is very close to the tunnel one.

A reason for the phonon singularities being shifted towards higher energies may be the complex structure of the contact. Since the control over the thickness of the deposited aluminum layer was not precise enough, its thickness probably proved to be greater than the conventional value. The aluminum film was not oxidized to the full depth during the subsequent treatment in peroxide. This mechanism may be an explanation for the peaks on the EPI function of the contact in Fig.\hyperref[Fig8]{8d-f} being smeared and shifted. There, the thickness of the aluminum film was smaller, and the surface of the niobium electrode was affected in the process of its oxidation. Given the conductivity of lower niobium oxides, it may be assumed that they affect the spectrum. However, it is not possible to accurately assess the structure of the contact. From a practical point of view, we have a resistance connected in series that shifts the phonon peaks in the spectrum to the region of higher energies, which can be taken into account if the true position of the phonon features is known.

Figure \hyperref[Fig9]{9} shows all three EPI functions reconstructed from the second derivatives of the $I-V$ curve in the superconducting state in comparison with the EPI functions, one of which was reconstructed from tunnel data \cite{28} and the other from the second derivative of the $I-V$ curve of the point contact in the normal condition \cite{24, 25}. The quality is apparently the best for the third curve adjusted using the positions of the phonon peaks. It is in the case of the point contact corresponding to this curve that the volume of niobium near the constriction, in which the nonlinearity of the $I-V$ curve is formed, is the most perfect.

\subsection{Point contacts based on $MgB_2$}
We have considered thus far conventional, virtually isotropic metals with a sufficiently long coherence length and a single gap. However, most of the known superconductors feature strong anisotropy, much smaller anisotropic coherence lengths, and the presence in some cases of two super­conducting gaps. It is a challenging problem to predict how the EPI will manifest itself in superconducting point contacts based on them; however, we use the previously developed techniques to reconstruct EPI functions for such materials.

One such object is $MgB_2$ \cite{17}, which has two superconducting gaps in the range $\Delta\sim2\div7~2\div7$~meV. It should be stressed that the value of the larger gap, 7~meV, refers to point contact measurements. The maximum gap amplitude in a number of tunneling experiments is as high as $\sim10\div11$~meV \cite{29,30,31}. The constant $\rho l=2.4\cdot10^{-12}\text{Ohm}\cdot\text{cm}^2$ differs by a factor slightly more than two from that of tantalum; therefore, given that resistance values are close to each other, the diameters of the obtained point contacts are slightly larger.
Contacts (Fig.\hyperref[Fig10]{10a}) were created by touching the $MgB_2$ film with a bulk silver counter-electrode \cite{32,33,34}. The film was highly oriented along the $c$-axis directed perpendicular to the surface of the $Al_2O_3$ substrate. At the same time, the crystal structure of the film in the $ab$ plane was randomly oriented.

The temperature of the superconducting transition of the film is $39~K$, and the transition width is $\sim0.2~K$. The resistance ratio $\rho(300~K)/\rho(40~K)\sim2.3$ \cite{35}, and the specific resistance of the sample at room temperature corresponded to polycrystalline material. The elastic mean free path in the film at a liquid helium temperature is of the order of 4~nm, the coherence length $\xi_0$=12~nm is almost an order of magnitude smaller than that in tantalum, and, given the elastic relaxation length, the reduced coherence length in the unperturbed material is only $\zeta\sim 3$~nm. The resistance of the point contacts ranged from 36~$\Omega$ to 150~$\Omega$, which, if evaluated using the Sharvin formula, yields diameters $d\sim2.2\div4.4$~nm. This can provide in the ideal case a mode of electron passage close to the ballistic one. For point contacts similar to those shown in Fig.\hyperref[Fig10]{10}, weakly pronounced, strongly smeared singularities are observed in the normal state in the region of phonon energies with a high background level \cite{32}. Thus, the actual regime of the passage of electrons is still not ballistic, and the given estimate of the diameters is an estimate from below.

In plotting the superconducting background curves, we were guided by the sum rule, indirectly taking into account the assumed shape of the absent normal-state curves; therefore, the background curves are not completely monotonic. Given the above, it may be assumed that for these contacts, as well as for the niobium contact considered last, only the center near the constriction is directly contaminated, albeit to a somewhat lesser extent, since small bends are observed in the background curves in the vicinity of phonon energies. The remaining volume of the point-contact and the banks preserve their initial structure and purity. Since the elastic path length in the banks is relatively small and the coherence length is also small, no selection of phonons with small group velocities similar to that in tantalum is observed here.

Figure \hyperref[Fig10]{10b} shows the EPI functions reconstructed on the basis of second derivatives after background curves are subtracted. Despite  some  differen ces, these functions are very close to each other.

Finally, Fig.\hyperref[Fig11]{11} displays the EPI function obtained by averaging over these five contacts in comparison with the point-contact functions for $ZrB_2$, which belongs to the same crystallographic group \cite{33}. It can be seen that the functions are rather similar in shape and position of phonon singularities, which are located for the averaged spectrum at energies $3\div5$~meV lower than those for $ZrB_2$, with the exception of the position of the first maximum, which is shifted to higher energies by $\sim16$~meV. This phenomenon is quite understandable, since this mode is associated withvibrations of $Zr$ ions whose atomic mass is $\sim3.75$ larger than that of $Mg$ ions.

An overall conclusion may be drawn that the method applied enables a fairly accurate reconstruction of the shape of the EPI function on the basis of superconducting point­contact spectra.

\subsection{Point contacts based on $NbSe_2$}

The compound $2H-NbSe_2$ is much more challenging for point-contact research than $MgB_2$. This layered and easily splittable superconductor is highly anisotropic. The specific resistance of this compound (${{\rho }_{ab}}\sim 2\cdot{{10}^{-4}}\Omega\cdot$cm; ${{\rho}_{c}}\sim{{10}^{-3}}\Omega\cdot$cm \cite{36}) and the value of the constant $\rho {{l}_{ab}}=2.2\cdot {{10}^{-11}}\Omega\cdot \text{cm}^{2}$  are more than an order of magnitude larger than those for conventional metals. Therefore, with comparable dimensions, the resistance values of the created point contacts are high by the 'standards' of conventional PC spectroscopy. The elastic path length in the $ab$ plane was approximately 33~nm for an unperturbed material at a temperature close to $T_C$. The coherence length in this material is even smaller than in $MgB_2$: $\xi_{ab}(0)=7.8$~nm, $\xi_{c}(0)=2.6$~nm. Given the elastic relaxation length, the reduced coherence length in the unperturbed material in the $ab$ plane is $\zeta\sim6.3$~nm. There are also two superconducting gaps in the range $\Delta\sim1.1\div2.5$~meV \cite{17}. In addition, noise in point-contact spectra strongly depends on the quality of seemingly very similar single crystals. Thus, for example, point contacts formed on the basis of a single crystal with $\rho_{300}/\rho_{res}\sim30$ featured a very low noise level at resistance values up to 1~k$\Omega$, while the noise level in single crystals with $\rho_{300}/\rho_{res}\sim100$ at resistance values above 10~$\Omega$ was very high. A possible reason for this phenomenon is the incommensurate charge density waves (CDWs) that exist in this compound. In the first case, single crystals are not very perfect, and there are effective pinning centers that hinder the CDW motion. In the second case, these centers are few, and the motion of the CDW in the vicinity of the contact in the current-passage state generates such noises.
\begin{figure}[]
\includegraphics[width=8.3cm,angle=0]{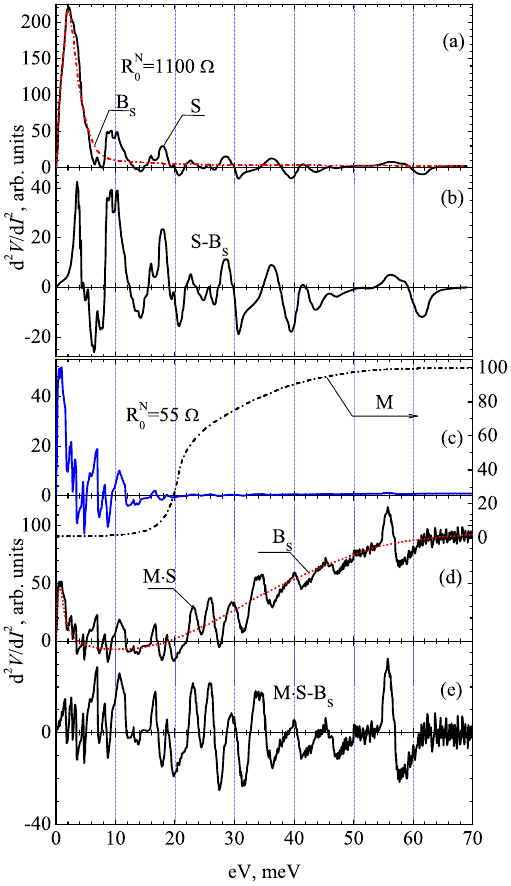}
\caption{Curves (a,b) for point contacts $S-c-N$ ($NbSe_{2}-Cu$) $R_{0}^{N}$=1100~$\Omega$; (c-e) point contacts $S-c-S$ ($NbSe_{2}-NbSe_2$) $R_{0}^{N}$=55~$\Omega$, T=4.2~K, H=0. (a) Second derivative of the curve and the background curve, (b) difference curve. (c) Second derivative of the $I-V$ curve and the scale curve, (d) amplitude-adjusted second derivative of the curve and the background curve, (e) difference curve.}
\label{Fig12}
\end{figure}
\begin{figure}[]
\includegraphics[width=8.3cm,angle=0]{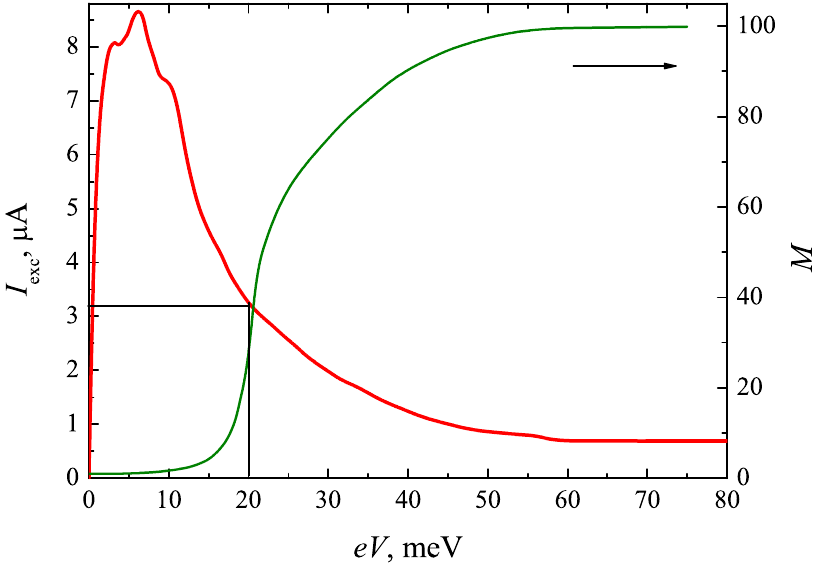}
\caption{Excess current of the $NbSe_{2}-NbSe_2$ point contact as a function of bias and the scale curve $M$ (Fig.\hyperref[Fig12]{12c-e}). The segments show the values of the excess current and voltage that correspond to the onset of the sector of growth on the scale curve.}
\label{Fig13}
\end{figure}

The single crystals under study had the shape of plates several millimeters in size and about 0.1~mm thick with natural faceting. The axis of the created point contacts was parallel to the $ab$ plane. In producing point contacts, the contact area was cut off with a blade, since the faces of natural growth of single crystals featured poor superconducting properties. Therefore, the crystal structure near the cut was highly disturbed.

\begin{figure}[]
\includegraphics[width=8.3cm,angle=0]{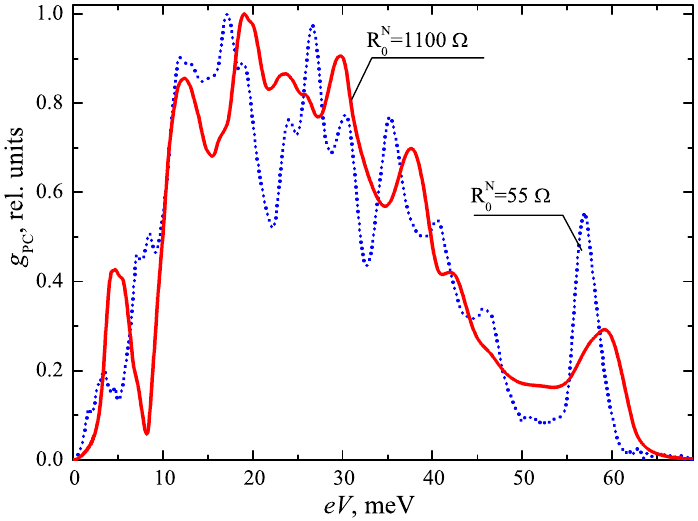}
\caption{EPI functions of $2H-NbSe_2$ normalized to one, reconstructed from second derivatives of the $I-V$ curve displayed in Fig.\hyperref[Fig12]{12} after subtracting the superconducting background.}
\label{Fig14}
\end{figure}

We present here the results of a study of two point contacts based on $2H-NbSe_2$ with different resistances. The diameter of the first $NbSe_{2}-Cu$ heterocontact (Fig.\hyperref[Fig12]{12a,b}) with a resistance of about 1.1~k$\Omega$ is close to the diameters of the $MgB_2$-based point contacts considered above and, if estimated using the Sharvin formula, is $d\approx2$~nm \cite{16,17}. Consequently, the reduced coherence length is in this case approximately three times the diameter. The diameter of the second homocontact (Figs\hyperref[Fig12]{12c-e}) with a resistance of about 55~$\Omega$ is much larger, d$\approx$17~nm \cite{16,37}. Thus, the diameter of this contact is already almost three times the reduced coherence length. Nevertheless, the elastic relaxation length for the undisturbed material is in both cases greater than the diameters of the contacts.

None of the point contacts studied by us \cite{16} had in the normal state phononrelated singularities on the second derivatives of the $I-V$ curve. In the transition of point contacts to the superconducting state, the second derivatives of the $I-V$ curve exhibit singularities that extend up to energies of about 65~meV.

Hence, it may be concluded that we are dealing in this case with point contacts in which defects do not occupy their entire volume, but rather are concen­trated near the constriction. Since the conversion of Andreev electrons into Cooper pairs occurs on the reduced coherence length, inelastic scattering of nonequilibrium phonons by Andreev electrons occurs for a point contact with a large diameter in the region of strong current density. However, Andreev electrons are not "smeared" here over a large volume, as is the case in ballistic contacts, in agreement with the predictions of the theory, but are concentrated in the same region of high current density. Therefore, as the bias grows, the density of nonequilibrium phonons increases, which leads to a rapid suppression of the excess current at biases above 15~meV in this point contact.

The intensity of the spectrum also begins to decrease following a decrease in the excess current. The situation here is in a sense similar to that observed in tantalum point contacts: there is a certain range of excess current values, in which the spectrum intensity changes insignificantly. Applied to this contact, this implies that the onset of the decrease in the spectrum intensity follows with a certain delay the onset of the decrease in the excess current.

Figure\hyperref[Fig13]{13} shows the excess current as a function of bias for this contact and the scale factor that adjusts the amplitude of the second derivative of the $I-V$ curve. The vertical and horizontal lines show the voltage that corresponds to the segment of a sharp increase in the scale curve and the corresponding value of the excess current. It is clearly seen that the decrease in the excess current occurs much more smoothly and significantly faster than the decrease in the amplitude of the second derivative. Therefore, the scale curve has a stepped shape similar to tantalum point contacts. This curve is scaled in such a way that $M_{min}$=1 and $M_{max}$=100.

Figure\hyperref[Fig14]{14} shows the EPI functions reconstructed for these contacts. There is good agreement between these functions reconstructed from such differing point contacts, especially given the high lability of the point contact spectra of $NbSe_2$ \cite{16}.

We now consider in more detail the low-resistance point contact. Figure\hyperref[Fig13]{13} shows that there is no critical current in this homocontact, and the excess current is only 8~$\mu$A, a value which is well below the dirty limit. Nevertheless, the broadening or smearing of the EPI function, characteristic of a material with a broken crystal lattice, is not observed for this contact. Thermal broadening of the phonon peaks is not observed either, indicating that the suppression of the excess current with an increase in the contact bias occurs due to other reasons. The small value of the excess current, which is due to the inhomogeneity of the point contact, is associated with the fact that the violation of the crystal structure results for $NbSe_2$, as for many other superconducting compounds, in the suppression of superconductivity. Part of the contact volume is filled with a nonsuperconducting material with a broken crystal lattice. Owing to this, the EPI spectrum is formed in that part of the point contact where the crystal structure of the material is not disturbed, which provides sharp and clear phonon peaks. As regards the magnitude of the excess current, the available formulas assume the presence of a homogeneous superconductor throughout the entire volume.

\section{Elastic point­contact spectroscopy}
In reconstructing the spectral EPI function $g(\omega)$ in elastic Rowell-McMillan spectroscopy, the normalized differential conductivity of superconducting tunneling junctions is used, which is referred to as the density of tunneling states. The latter does not contain this function in an explicit form; to reconstruct it, a numerical solution of Eliashberg's integral equations of the phonon theory of superconductivity is used.

\begin{table*}[htbp]
\caption[]{Estimated elastic contribution to the spectrum of some superconductors $\delta_{rel}$ normalized to the contribution in lead and the values of the energy gap and temperature of the superconducting transition}
\renewcommand{\tabcolsep}{0.4cm}
\begin{tabular}{|c|c|c|c|c|c|c|c|c|} \hline
Superconductor &	$Pb$ & $Nb$ & $In$ &	$Sn$ & $Ta$ & $Al$ & $NbSe_2$ & $MgB_2$ \\ \hline
$\delta_{rel}$ &	1	& 0.21	&0.21	& 0.078	& 0.063	& $1.68\cdot10^{-3}$ &0.023 & 0.24 \\ \hline
$\Delta_0$,~meV	& 1.365	& 1.6	&0.525 &	0.575&	0.7&	0.17&	$1.07\div2.48$ & $1.8\div7.4$ \\ \hline
$T_C$,~K &	7.2&	9.2&	3.42&	3.722&	4.47&	1.181&	7.2	&39 \\ \hline
\end{tabular}
\label{Table2}
\end{table*}

If the elastic component of the current is taken into account for superconducting point contacts with direct conductivity, a term additional to the $I-V$ curve nonlinearities described by Eqn\ref{eq__1} emerges. This additional nonlinearity, which is associated with the dependence of the superconducting gap on the bias at the contact, arises due to the electron-phonon renormalization of the superconductor energy spectrum. Such nonlinearity exists for all point contacts, but its value is very sensitive to the strength of the electron-phonon interaction, and it can differ for various superconductors by a factor of several hundred or even thousands. The estimate of the expected elastic contribution to the spectrum is proportional to $\sim(T_{C}/\theta_{D})^2$, where $\theta_D$ is the Debye temperature. Lead features one of the largest elastic contributions to the spectrum. Table\ref{Table2} shows the elastic contributions of some superconductors in comparison with that of lead, together with the values of superconducting gaps and critical temperatures.

The normalized differential conductivity of point contacts with direct conductivity differs from that for tunnel junctions due to the presence of excess current associated with Andreev reflection processes. The corresponding formulas for these conductivities for point and tunnel contacts are given in \cite{19}.

We now consider more closely the elastic contribution for pointlike point contacts. Elastic processes were shown in \cite{7} to result in the emergence of \emph{maxima of differential} \emph{\textbf{conductivity}} in the region of characteristic phonon frequencies on the first derivative of the excess current. In contrast to the plot shown in Fig.\hyperref[Fig1]{1b)}, the single phonon line will manifest itself as a stepwise increase in the excess current. However, inelastic processes manifest themselves as \emph{maxima of the differential} \emph{\textbf{resistance}} on the first derivative of the excess current. Thus, the effects of these contributions are opposite to each other, and if amplitudes of these contributions are close to each other, they can be mutually reduced. It is very difficult to estimate in advance which contribution will prevail, since the available formulas are derived for the ballistic regime of point contacts.

At the same time, as was found for tantalum-based point contacts, the not very accurate fulfillment of the ballistic condition regarding the coherence length greatly increases the proportion of the inelastic contribution. Since the coherence length for many superconductors is often not too large in comparison with the diameters of point contacts, the predominance of the inelastic contribution is expected to be more likely. Anyway, this is exactly the case for the examples considered above. However, it was found for at least two superconductors, namely lead and indium, that the elastic contribution predominates in the spectra of point contacts with direct conductivity based on these materials. The same conclusion is undoubtedly true for mercury-based point contacts.

Although study \cite{7} does not contain direct indications that the EPI function is present in an explicit form in the spectra of point contacts with a predominant elastic contribution, the authors of \cite{18} used the conclusion about the manifestation of phonon peaks in the form of maxima of differential conductivity to successfully reconstruct these functions using lead and indium spectra. To reconstruct the EPI function from the elastic contribution to the spectrum of point contacts, techniques have been used that were successful in inelastic Andreev spectroscopy.

\subsection{Lead­based point contacts}
\begin{figure*}[tb]
\includegraphics[width=16.5cm,angle=0]{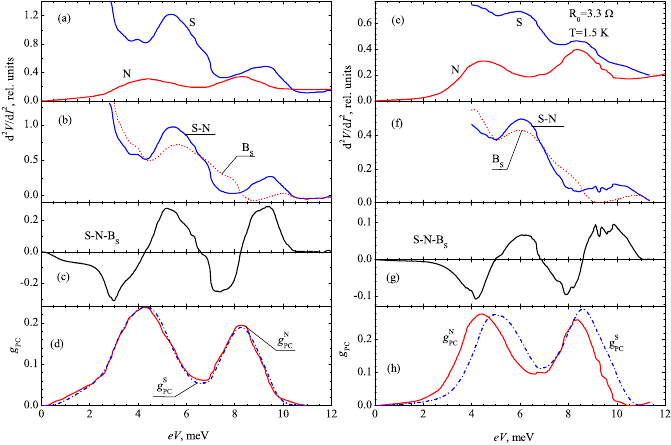}
\caption{Curves for $S-c-N$ $(Pb-Ru)$ (a-d) and $S-c-S$ $(Pb-Pb)$ (e-h) point contacts. (a,e) EPI spectra of a point contact in the normal ($N$) and superconducting ($S$) state, respectively; superconductivity is suppressed by a magnetic field. (b,f) Difference between the superconducting and normal spectra and the assumed shape of the background curve. (c, g) Difference curve (after background subtraction). (d, h) Point-contact EPI functions $g_{pc}^N$ and $g_{pc}^S$, the former function reconstructed from the normal state, and the latter by integrating the $S-N-B_S$ curve shown in (c, g)}.
\label{Fig15}
\end{figure*}

Lead has the constant $\lambda$=1.55, a value which is one of the highest among all elements. Mercury alone has a greater value, $\lambda$=1.6. Figure \hyperref[Fig15]{15a,e} shows the spectra of the $Pb-Ru$ heterocontact in the superconducting ($S$) and normal ($N$) states. The ruthenium EPI spectrum is much harder than that of lead and does not overlap with it in energy, so it is not shown in the figure. The spectrum intensity in the normal state is close to the maximum for a symmetric heterocontact; therefore, the contact can be considered a ballistic contact.

Figure \hyperref[Fig15]{15b,f} shows the difference curve and the assumed curve of the superconducting background. In constructing the background curve, we were guided by the same \emph{sum rule} as in the case of an inelastic contribution. As can be seen from the figure, the shape of the background curves for the elastic contribution radically differs from that for inelastic contributions. Energy dependence is described by a complex non­monotonic rather than smooth curve. This circumstance, unfortunately, allows a great deal of arbitrariness in the choice of the shape of the background curve and thus the shape of the reconstructed EPI function.

Nevertheless, as follows from the examples given below (see also \cite{18}), the deviations in shape from the EPI functions reconstructed from the spectra in the normal state are relatively small. They may well be due to the nonequivalence of volumes, from which information about the electron­phonon interaction is obtained, since these volumes differ by the degree of perfection of the crystal lattice.

Figure \hyperref[Fig15]{15c,g} shows the resulting curve after subtracting the superconducting background. As noted above, it follows from theory \cite{7} that elastic processes result in the emergence of maxima of differential conductivity in the vicinity of phonon peaks. Therefore, the curve obtained as a result of integration is inverted (Fig.\hyperref[Fig15]{15d,h}). The shape of this curve is seen to virtually coincide with that of the EPI function of lead reconstructed from the spectrum in the normal state. Thus, there is no shift of the phonon peaks in the EPI function reconstructed from the elastic addition to the $S-c-N$ contact compared with the EPI function for the normal state.

Figure \hyperref[Fig15]{15e-h} displays data for the homocontact of the $Pb-Pb$ contact. The curves are processed here in the same way as in the case shown in Fig.\hyperref[Fig15]{15a-d}. A difference from the heterocontact should be immediately noted: the phonon peaks on the reconstructed curve are shifted here to the region of higher energies by an amount of the order of $\Delta$.

Despite the difference between tunneling and point contacts, the authors of \cite{19} were the first to apply, in addition to the empirical method for reconstruction of the EPI function described above, the iterative Rowell-McMillan method for solving the Eliashberg equations in reconstructing the EPI function of lead. They used to this end experimental data for point contacts with direct conductivity and the values of the real and imaginary parts of $\Delta(\omega)$ from the table in \cite{38}.

\subsection{Indium­based point contacts}
\begin{figure}[]
\includegraphics[width=8.3cm,angle=0]{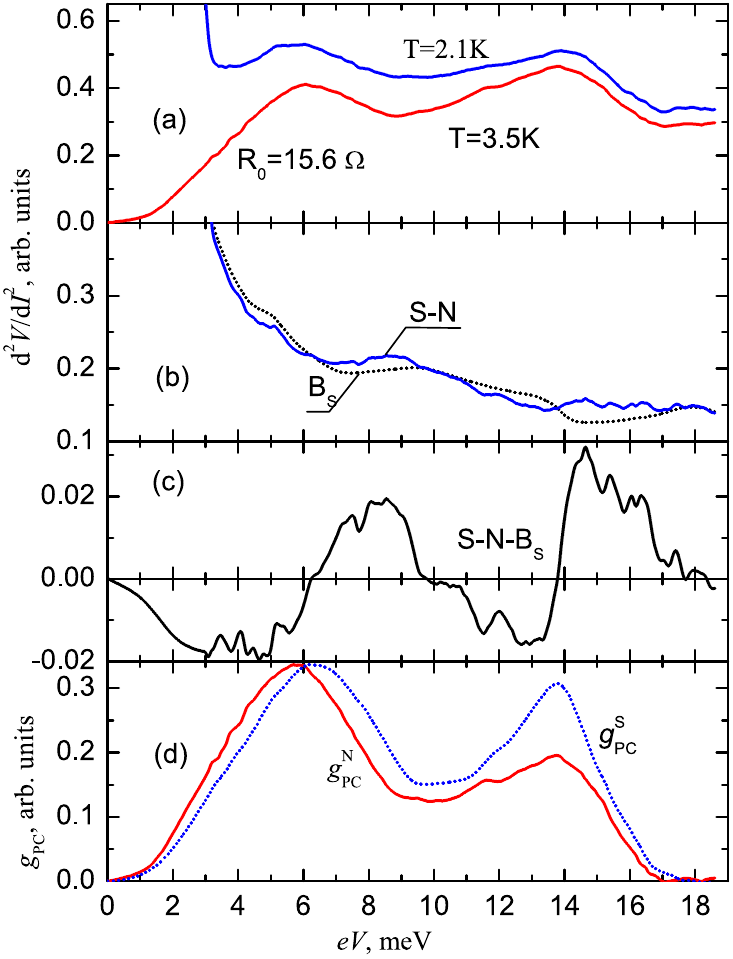}
\caption{(a) EPI spectra of an $In-In$ point contact in the normal and superconducting states at various temperatures. (b) Differences between the superconducting and normal spectra and the assumed shape of the background curve at $T=2.1$~K, $T/T_{C}=0.62$ and $\Delta=0.89\Delta_0$. (c) Differential curve (with the background subtracted.) (d) Point contact EPI function $g_{pc}^S$ reconstructed by integrating the $S-N-B_S$ curve shown in fig. c in comparison with the EPI function $g_{pc}^N$ of the normal state.}
\label{Fig16}
\end{figure}

Table\ref{Table2} shows that the elastic addition to the spectrum in indium is almost 5~times smaller than that in lead, but at the same time is 2.7~times larger than that in tin. Since the temperature of the superconducting transition and the gap in indium are only $\sim$8\% lower than in tin, the inelastic superconducting contributions to the spectra in these metals should be very close to each other. Thus, given the opposite effect of the elastic and inelastic contributions, it may be expected that the resulting elastic contribution in indium is significantly weakened.
Figure \hyperref[Fig16]{16} shows the spectra of an indium homocontact in the normal and superconducting states and the results of processing these spectra in reconstructing the EPI function from an elastic addition \cite{18}. As in the case of lead homocontacts, a shift of the phonon peaks in the EPI function towards higher energies is observed by an amount of the order of the energy gap. The elastic addition here, as expected, is very small in comparison with that for lead.

It should be noted that the manifestation of elastic or inelastic contributions in the spectra of superconducting point contacts depends not on the absolute value of these contributions but rather on the relationship between them. For example, as follows from Table\ref{Table2}, the elastic contribution to the spectrum for $Nb$ coincides with that for $In$, and even exceeds it for $MgB_2$. However, since the inelastic contribution is proportional to the excess current determined by the energy gap, which is much smaller in $In$, the resulting contribution for $Nb$ and $MgB_2$ turns out to be inelastic. Nevertheless, the amount of excess current can be reduced by creating point contacts with any transparency of the tunneling barrier between the electrodes. A decrease in the barrier transparency will eventually lead to the predominance of the elastic contribution.

\section{Discussion of results}
The theories of both elastic and inelastic Andreev EPI spectroscopy refer to the ballistic regime of the passage of electrons through a constriction. A theory for the diffusive limit is presented in Ref.\cite{39}, but it refers to a long channel with dirty banks and a jumper and is not suitable for describing the cases most often encountered in experiments. There is no theoretical model for the inhomogeneous contacts discussed in Section 5. Nevertheless, the approach used for ballistic point contacts turned out to be applicable with some reservations to these objects as well. If the nondissipative diffusive regime of electron transit through the defect region in the constriction is operative for spatially inhomogeneous point contacts, i.e., duplication of electrons persists, the second derivative of the $I-V$ curve of such contacts is largely similar to the difference between the second derivatives of the $I-V$ curve of the superconducting and normal states of ballistic contacts. Experiments with niobium point contacts show that even the dissipative mode of electron passage through the defect region can be used in some cases to reconstruct the EPI function, taking into account the emerging resistance connected in series.

The question may always arise concerning materials with an unknown EPI spectrum: is there dissipation? Statistics can apparently be helpful in this case. For example, the EPI function was unknown for $NbSe_{2}$. The point contacts used to reconstruct this function were radically different both in structure (homo-and heterocontacts) and resistance (55~$\Omega$ and 1100~$\Omega$, respectively). In addition, data are presented in \cite{17} for one more heterocontact with a resistance of 750~$\Omega$. The positions of the phonon singularities on all these EPI functions virtually coincide, which rules out the presence of a resistance connected in series.

There is also ambiguity in plotting the curves of the superconducting background and in choosing the position and amplitude of the step of the adjusting curve, which ensures the fulfillment of the sum rule for spectra with suppressed excess current. The behavior of the superconducting background usually corresponds at large biases to the general shape of the spectrum, while in the low-voltage region, the main requirement is to prevent the emergence of any artifact singularities on the difference curves in moving to the energy region of the gap or nonequilibrium (for tantalum point contacts) singularity. The missing section of the curve that begins at zero should also satisfy the same requirements. This implies that this segment of the curve should smoothly match its higher energy counterpart. In addition, the fulfillment of the sum rule can be improved by varying within narrow limits the shape of this part of the background curve.

Although the reason for the decrease in the amplitude of the difference curve (or simply the second derivative of the $I-V$ curve for inhomogeneous contacts) is a decrease in the excess current with increasing voltage, there is no linear relationship between these phenomena. In a certain range of variation of the excess current, the value of the superconducting addition to the spectrum initially depends weakly on the magnitude of the excess current, and then decreases rather sharply. Possible reasons for this behavior are discussed at the end of Subsection 4.2. This implies from a practical point of view that there is no need to monitor the dependence of excess current on voltage. It is sufficient to only monitor the amplitude of the curve that requires adjustment. If for known materials such as tantalum and niobium this curve is integrated before adjustment (see Fig.\hyperref[Fig6]{6}), its shape alone is sufficient to determine by means of straightforward selection the position and next the amplitude of the step of the adjusting curve. The choice of the position and amplitude of the step of the adjustment curve is not this obvious in the case of unknown materials with a complex spectrum. Only the experience in handling $NbSe_{2}$ can be referred to here. Figure \hyperref[Fig12]{12c-e} shows the second derivative of the $I-V$ curve that requires adjustment of the amplitude. The step position and amplitude were selected in this case in such a way that the adjusted curve had approximately the same amplitude in the entire energy range without sharp surges.

Such a double adjustment as background subtraction and compensation for the decrease in the amplitude of the curve in the high-energy region undoubtedly makes the results obtained more uncertain. A different matter is whether these variations are critically significant. If all the EPI functions reconstructed from the superconducting addition for tantalum homo-and heterocontacts are compared with each other (see review \cite{22}), their variations do not exceed the variations of the EPI functions reconstructed from the normal state. As for $NbSe_{2}$, the uncertainty is in this case significantly larger, resulting in variations in the shape of the reconstructed EPI function. However, the positions of the phonon singularities in this function are at least virtually independent of the position of the step and the amplitude of the adjusting curve.

An additional complicating factor may be the emergence in the derivatives of the $I-V$ curve of nonspectral singularities associated with the breakdown of superconductivity in the near-contact region. Such singularities may be of various natures, for example, thermal or associated with nonequilibrium processes. These singularities are not reproducible; their position depends on the temperature and the resistance of the point contact. Therefore, to exclude such singularities from consideration, certain statistics of the spectra are required.

Experiments show that it is much easier to create inhomogeneous contacts suitable for spectroscopy if superconductors with a covalent bond between atoms are used. Their crystal lattice is significantly stiffer than that of metal­bonded superconductors. If point contacts are produced employing mechanical methods, lattice distortions extend to a smaller distance from the constriction, which ensures a minimum volume of defective material.

The examples considered above do not encompass the entire variety of point contacts with regard to the ratio between their sizes and coherence lengths. Nevertheless, the limits of applicability of Andreev spectroscopy are already quite apparent. First, they are ballistic point contacts of superconductors with high critical parameters. It is very difficult to make such point contacts to transit to the normal state at low temperatures, while, in the superconducting state, the contributions to the $I-V$ curve nonlinearity from backscattering and nonequilibrium phonon scattering by Andreev electrons are comparable due to the small coherence length in such materials. It is almost impossible to separate these contributions. Also beyond the limits of applicability of the method are materials with a very small coherence length and relatively low critical current density, for example, the aforementioned $2H-NbSe_2$, but only if the contact axis is directed perpendicular to the layers. A stepped structure with discrete values of the differential resistance is observed on the current-voltage characteristics for such contacts (see Fig.~1 in \cite{40}). These steps are associated with the advancement of phase-slip surfaces into the bulk of the superconductor (see Fig.~2 in Ref. \cite{40}). Since such jumps extend to significant biases, it is apparently impossible to observe the EPI in this range. The situation is worse for materials with even shorter coherence lengths. However, a technique used for niobium point contacts seems to be applicable to such materials. If a normal-metal film is deposited on the surface of such materials, then, when a point contact is created, the constriction will be moved away from the superconductor by a sufficiently large distance so that the density of the current does not exceed the critical value owing to its spreading. In addition, as the constriction moves away from the super­conductor, the contribution to the nonlinearity of the $I-V$ curve from backscattering rapidly decreases. At the same time, the contribution to the nonlinearity of the $I-V$ curve of the point contact from the deposited film can be disregarded.

The effect of a magnetic field on the Andreev EPI spectra has been insufficiently studied. We are not aware of such experiments for ballistic point contacts that fully meet the requirements of the theory. As for the tantalum-based ballistic point contact, whose spectra significantly deviate from the theoretical predictions, a very strong effect of rather weak (several ten oersteds) magnetic fields (see Fig.~2 in \cite{14}) on the shape of the second derivative of the $I-V$ curve is observed. The effect of a magnetic field on the EPI spectra of type II superconductors was studied in the $ErNi_{2}B_{2}C$ point contact \cite{37}. A strong effect on the intensity and position of phonon singularities in the spectra has been found. However, since, unfortunately, too large a step in varying the magnetic field was used in the measurements, it is difficult to draw any quantitative conclusions from the data obtained. Nevertheless, a preliminary conclusion can be made that the Andreev EPI spectra should preferably be studied in the absence of a magnetic field.
\section{Conclusion}
\begin{enumerate}
\item To reconstruct the EPI function using superconducting point contacts, the second derivative of the $I-V$ curve should be used due to the presence of a superconducting background.
\item For ballistic point contacts created on the basis of superconductors with a large coherence length ($Al$, $Sn$), there is full agreement with the conclusions of theoretical studies. In subtracting the superconducting background, it is necessary to be guided by the sum rule.
\item For ballistic point contacts created on the basis of superconductors with a smaller coherence length ($Ta$), there is significant disagreement with theoretical predictions in the transition to the superconducting state. This deviation is due to the nonlinearity of the I-V curve of point contacts if the superconducting state is formed not only in a volume of the order of the contact diameter, as in the previous case, but also in the near-contact region, whose size is of the order of the reduced coherence length.
\item For inhomogeneous point contacts, a basic difference is observed between the second derivatives of the $I-V$ curve in the normal and superconducting states. In the normal state, the singularities associated with phonons are either absent or sharply weakened in the derivatives. In the superconducting state, singularities emerge on the second derivatives of the $I-V$ curve that enable the EPI functions to be reconstructed.
\item A variation of the excess current within a sufficiently wide range insignificantly affects the magnitude of the superconducting addition to the spectrum; however, if it decreases below a certain value, the amplitude of the addition is abruptly suppressed. The functional dependence of the decrease in the excess current on the bias at the contact does not coincide in this case with the dependence of the decrease in the intensity of the addition. The latter always decreases in a stepwise manner. The sum rule for contacts with suppressed excess current fails, which requires their spectra to be adjusted to reconstruct the EPI function.
\item For point contacts with direct conductivity, the superconducting elastic addition to the spectrum operates oppositely to the inelastic one; therefore, the addition whose contribution prevails is manifested as a result of superposition. The positions of the phonon peaks in the EPI function reconstructed based on the elastic addition coincide for $S-c-N$ type point contacts with those in the normal state, while, for $S-c-S$ contacts, it is shifted to the high-energy region by the value of $\Delta$.
\end{enumerate}

This study was financially supported by the National Academy of Sciences of Ukraine under project FTs 4-19.


\end{document}